\documentclass[letterpaper,twocolumn,10pt]{article}
\newif\ifarxiv
\arxivfalse 

\ifarxiv
\else
\usepackage{usenix2019_v3}
\fi

\newif\ifextended
\extendedtrue

\ifextended
\else
\fi

\usepackage{latexsym,amssymb,stmaryrd}
\usepackage{amsmath}
\usepackage{amsthm}
\usepackage{colortbl}
\usepackage{color} 
\usepackage{graphics}
\usepackage{graphicx}
\usepackage{tikz}
\usepackage{multicol}
\usepackage{multirow}
\usepackage{url}
\usepackage{arydshln}
\usepackage{pgfplots}
\usepackage{listings}
\usepackage{todo}
\usepackage{longtable}

\usepackage{enumitem}

\ifarxiv
\else
\usepgfplotslibrary{external}
\fi

\usepackage{ctable}
\lstdefinelanguage
   [x64]{Assembler}     
   [x86masm]{Assembler} 
   {morekeywords={CDQE,CQO,CMPSQ,CMPXCHG16B,JRCXZ,LODSQ,MOVSXD, %
                  POPFQ,PUSHFQ,SCASQ,STOSQ,IRETQ,RDTSCP,SWAPGS, %
                  rax,rdx,rcx,rbx,rsi,rdi,rsp,rbp, %
                  r8,r8d,r8w,r8b,r9,r9d,r9w,r9b, %
                  r10,r10d,r10w,r10b,r11,r11d,r11w,r11b, %
                  r12,r12d,r12w,r12b,r13,r13d,r13w,r13b, %
                  r14,r14d,r14w,r14b,r15,r15d,r15w,r15b,rip},
     columns=fixed,
     basicstyle=\small\ttfamily\upshape,
   } 

\theoremstyle{definition}

\newtheorem{example}{Example}

\newcommand{\address}{\ensuremath{\mathbb{A}}}
\newcommand{\register}{\ensuremath{\mathbb{R}}}
\newcommand{\operand}{\ensuremath{\mathbb{O}}}
\newcommand{\str}{\ensuremath{\mathbb{S}}}
\newcommand{\num}{\ensuremath{\mathbb{Z}_{64}}}

\newcommand{\ddisasm}{\ifanonymous\texttt{XXXX}\else\texttt{Ddisasm}\fi}
\newcommand{\gtirb}{\ifanonymous\texttt{YYYY}\else GTIRB\fi}
\newcommand{\Ramblr}{\texttt{Ramblr}}

\lstMakeShortInline|
\newcounter{rcount}
\newcommand{\rcount}[1]{\refstepcounter{rcount}(\thercount)\label{#1}}

\newcounter{pcount}
\newcommand{\pcount}[1]{\refstepcounter{pcount}P\thepcount\label{#1}}

\newcommand{\ibi}{IBI}
\newcommand{\dap}{DAP}

\newcommand\pro{\item[$+$]}
\newcommand\con{\item[$-$]}

\newenvironment{drule}[1]{
  \def\myenvargumentII{#1}
  \noindent
  \begin{tabular*}{\linewidth}{@{}l@{\extracolsep{\fill}}r}
}
{
  & \rcount{\myenvargumentII}\\
  \end{tabular*}
}

\newif\ifsrpm
\srpmfalse 

\newif\ifanonymous
\anonymousfalse

\begin{document}

\title{Datalog Disassembly}
\ifanonymous
\else
\author{
  {\rm Antonio Flores-Montoya}\\
  GrammaTech, Inc.\\
  {\tt afloresmontoya@grammatech.com}
\and
  {\rm Eric Schulte}\\
  GrammaTech, Inc.\\
  {\tt eschulte@grammatech.com}
}
\fi

\maketitle

\begin{abstract}

  Disassembly is fundamental to binary analysis and rewriting.  We
  present a novel disassembly technique that takes a stripped binary
  and produces reassembleable assembly code.  The resulting assembly
  code has accurate symbolic information, providing cross-references
  for analysis and to enable adjustment of code and data pointers to
  accommodate rewriting.  Our technique features multiple static
  analyses and heuristics in a combined Datalog implementation.  We
  argue that Datalog's inference process is particularly well suited
  for disassembly and the required analyses.  Our implementation and
  experiments support this claim.  We have implemented our approach
  into an open-source tool called \ddisasm{}{\ifanonymous\footnote{The name of the tool
    has been substituted by \texttt{XXXX} for anonymization.}\fi}.  In extensive experiments
  in which we rewrite thousands of x64 binaries we find \ddisasm{} is
  both faster and more accurate than the current state-of-the-art
  binary reassembling tool, \Ramblr{}.
\end{abstract}

\section{Introduction}

Software is increasingly ubiquitous and the identification and
mitigation of software vulnerabilities is increasingly essential to
the functioning of modern society.  In many cases---e.g., COTS or
legacy binaries, libraries, and drivers---source code is not available
so identification and mitigation requires binary analysis and rewriting.  Many
disassemblers~\cite{gnubinaryutilities,ramblr,uroboros,superset,wartell2014shingled,probabilistic,speculative,Kruegel},
analysis
frameworks~\cite{ida,ghidra,binja,hopper,angr,bap,radare,macaw,pharos,b2r2},
rewriting
frameworks~\cite{reins,superset,irdb,mcsema,secondwrite,PSI,diablo,eel,bistro},
and reassembling tools~\cite{ramblr,uroboros,probabilistic} have been
developed to support this need.  Many applications depend on these
tools including binary hardening with control flow
protection~\cite{bincfi,typearmor,vci,opaquecfi,ccfir,pappas2012smashing},
memory protections~\cite{stackarmor,binarmor,prasad2003binary},
memory diversity~\cite{mvarmor,codearmor}, binary
refactoring~\cite{tilevich2005binary}, binary
instrumentation~\cite{romer1997instrumentation}, and binary
optimization~\cite{diablo,schwarz2001plto,romer1997instrumentation}.

Modifying a binary is not easy.  Machine code is not designed to be
modified and the compilation and assembly process discards essential
information.  In general reversing assembly is not decidable.  The
information required to produce reassembleable disassembly includes:

\begin{description}[noitemsep,nolistsep]
\item[Instruction boundaries] Recovering where instructions start and end
  can be challenging especially in architectures such as
  x64 that have variable length instructions, dense instruction
  sets\footnote{Almost any combination of
    bytes corresponds to a valid instruction.}, and sometimes
  interleave code and data. This problem is also referred as
  \emph{content classification}.
\item[Symbolization information] In binaries, there is no distinction
  between a number that represents a literal
  and a reference that points to a location in the code or data.
  If we modify a binary---e.g., by moving a block of
  code---all references pointing to that block, and to all of
  the subsequently shifted blocks, have to be updated. On the other
  hand, literals, even if they coincide with the address of a block,
  have to remain unchanged. This problem is also referred to as
  \emph{Literal Reference Disambiguation}.
\end{description}

We have developed a
disassembler that infers precise information for both
questions and thus generates reassembleable assembly for a
large variety of programs. These problems are not solvable in general
so our approach leverages a combination of static program analysis and
heuristics derived from empirical analysis of common compiler and
assembler idioms.  The static analysis, heuristics, and their
combination are implemented in Datalog. Datalog is a declarative
language that can be used to express dataflow analyses very
concisely~\cite{DatalogProgramAnalysis} and it has recently gained
attention with the appearance of engines such as
Souffle~\cite{souffle} that generate highly efficient parallel C++
code from a Datalog program. We argue that Datalog is so well suited
to the implementation of a disassembler that it represents a
qualitative change in what is possible in terms of accuracy and
efficiency.

We can conceptualize disassembly as taking a series of
decisions.
  {\em Instruction boundary identification} (IBI) amounts to deciding, for
    each address $x$ in an executable section, whether $x$ represent the
    beginning of an instruction or not.
    {\em Symbolization} amounts to deciding for each number that
    appears inside an instruction operand or data section whether it
    corresponds to a literal or to a symbolic expression and what kind
    of symbolic expression it is.\footnote{E.g.,
      \lstinline{symbol}, \lstinline{symbol+constant}, or \lstinline{symbol-symbol}.}

The high level approach for each of these decisions is the same.  A
variety of static analyses are performed that gather evidence for
possible interpretations.  Then, Datalog rules assign weights to the
evidence and aggregate the results for each interpretation.  Finally,
a decision is taken according to the aggregate weight of each possible
interpretation. Our implementation infers instruction boundaries first
(described in Sec.~\ref{sec:instructions}).  Then it performs
several static analyses to support the symbolization procedure: the
computation of def-use chains, a novel register value analysis, and a
data access pattern analysis described in Sec.~\ref{sec:def-use},
\ref{sec:value-analysis}, and \ref{sec:data-access} respectively.
Finally, it combines the results of the static analyses with other
heuristics to inform symbolization.  All these steps are
implemented in a single Datalog program. It is worth noting
that---Datalog being a purely declarative language---the sequence in
which
each of the disassembly steps is computed stems solely from the
logical dependencies among the different Datalog rules.
Combining multiple analyses and heuristics is essential
to achieve high accuracy for \ibi{} and symbolization.
No individual analysis or heuristic provides perfect information but
by combining several, \ddisasm{}
maximizes its chances to reach the right conclusion.
The declarative nature of Datalog makes this combination easy.

We have tested \ddisasm{} and compared it to \Ramblr{} \cite{ramblr} (the current best
published disassembler that produces reassembleable assembly) on 200
benchmark programs including 106 Coreutils, 25 real world
applications, and 69 binaries from DARPA's Cyber Grand Challenge
(CGC)~\cite{cgc}.
We compile each benchmark using $7$ compilers and $5$ or $6$
optimization flags (depending on the benchmark) yielding a total of
7658 unique binaries ($888$ MB of binary data).  We
compare the precision of the disassemblers by making
semantics-preserving modifications to the assembly code---we ``stretch'' the
program's code address space by adding NOPs at regular intervals---reassembling
the modified assembly code, and then running the test suites
distributed with the binaries to check that they retain functionality.
Additionally, we evaluate the symbolization step by
comparing the results of the disassembler to the ground truth
extracted from binaries generated with all relocation information.
Finally, we compare the disassemblers in terms of the time taken by
the disassembly process. \ddisasm{} is faster
and more accurate than \Ramblr{}.

Our contributions are:
\begin{enumerate}[noitemsep,nolistsep]
\item We present a new disassembly framework based on combining static
  analysis and heuristics expressed in Datalog.  This framework
  enables much faster development and empirical
  evaluation of new heuristics and analyses.

\item We present multiple static analyses implemented in this framework
  to support building reassembleable assembly.

\item We present multiple empirically motivated heuristics that are
  effective in inferring the necessary information to produce
  reassembleable assembly.

\item Our implementation is
  called \ddisasm{} and it is open source and publicly
  available\footnote{\ifanonymous \url{https://github.com/(anonymized)}\else\url{https://github.com/GrammaTech/ddisasm}\fi}. \ddisasm{}
  produces assembly text as well as an intermediate representation (IR)
  tailored for binary analysis and
  rewriting\footnote{\ifanonymous \url{https://github.com/(anonymized)}\else\url{https://github.com/GrammaTech/gtirb}\fi}.

\item We demonstrate the effectiveness of our approach through an
  extensive experimental evaluation of over 7658 binaries in which we
  compare \ddisasm{} to the state-of-the-art tool in reassembleable
  disassembly \Ramblr{}.
\end{enumerate}

\section{Related Work}



\subsection{Disassemblers}

Bin-CFI~\cite{bincfi} is an early work in reassembleable disassembly.
This work requires relocation information (avoiding the need for
symbolization).  With this information, disassembly is reduced to the
problem of \ibi{}.  Bin-CFI combines linear disasssembly with the
backward propagation of invalid opcodes and invalid jumps.  Our \ibi{}
also propagates invalid opcodes and jumps backwards, but it couples it
with a more sophisticated forward traversal.

Many other works focus solely on \ibi{} \cite{probabilistic,
  wartell2014shingled,speculative,Kruegel}.  None of these address
symbolization.  In general they try to obtain a superset
of all possible instructions or basic blocks in the binary and then
determine which ones are real using heuristics.
This idea is also present in our approach.  Both Miller et
al.~\cite{probabilistic} and Wartell et al.~\cite{wartell2014shingled}
use probabilistic methods to determine which addresses contain
instructions.  In the former, probabilistic techniques with weighted
heuristics are used to estimate the probability that each offset in
the code section is the start of an instruction. In the latter, a
probabilistic finite state machine is trained on a large corpus of
disassembled programs to learn common opcode operand pairs.  These
pairs are used to select among possible assembly codes.

Despite all the work on disassembly, there are disagreements on how
often challenging features for \ibi{}---e.g., overlapping
instructions, data in code sections, and multi-entry functions---are
present in real
code~\cite{probabilistic,meng2016binary,andriesse}. Our experience
matches \cite{andriesse} for GCC and Clang, in that we did not
find data in executable sections nor overlapping
instructions in ELF binaries. However, this is not true for
the Intel compiler (ICC) which often allocates jump tables in executable
sections.

There are only a few systems that address the symbolization problem
directly.  Uroboros~\cite{uroboros} uses {\em linear disassembly} as
introduced by Bin-CFI~\cite{bincfi} and adds heuristics for
symbolization.  The authors distinguish four classes of symbolization
depending on if the source and target of the reference are present in
code or data.  The difficulty of each class is assessed and partial
solutions are proposed for each class.

\Ramblr{}~\cite{ramblr} is the closest related work. It improves upon
Uroboros with increasingly sophisticated static analyses.  \Ramblr{}
is part of the Angr framework for binary analysis~\cite{angr}.  Our
system also uses static analyses in combination with heuristics.  Our
static analyses (Sec. \ref{sec:analysis}) are specially tailored to
enable symbolization while remaining efficient. 
Moreover, our Datalog
implementation allow us to easily combine analysis results and
heuristics.

RetroWrite~\cite{retrowrite} also performs symbolization, but only for
position independent code (PIC) as it relies on relocations.  In
Sec.~\ref{sec:symbolization-experiments}, we argue why we believe that
relocations are not enough to perform symbolization even for PIC.

\subsection{Rewriting Systems}
REINS~\cite{reins} rewrites binaries in such a way as to avoid making
difficult decisions about symbolization.  REINS partitions the memory
of rewritten programs into untrusted \emph{low-memory} which includes
rewritten code and trusted \emph{high-memory} (divided at a power of two
for efficient guarding).  They implement a lightweight binary lookup
table to rewrite each old jump targets with a tagged pointer to its
new location in the rewritten code.  REINS targets Windows binaries
and its main goal is to rewrite untrusted code to execute it
safely. REINS uses IDA Pro~\cite{idapro} to perform \ibi{} and to resolve indirect jumps.

SecondWrite \cite{secondwrite} also avoids making symbolization decisions
by translating jump targets at their point of usage. They do a conservative
identification of code and data by performing
speculative disassembly and keeping the original code section intact.
Any data in the code section can still be accessed, but jumps
and call targets are translated to a rewritten code section.
SecondWrite disassembles to LLVM IR.


MULTIVERSE~\cite{superset} goes a step further than SecondWrite
and also avoids making code location determinations by treating every possible
instruction offset as a valid instruction.  Similarly to SecondWrite, it avoids making
symbolization determinations by generating rewritten executables in
which every indirect control flow is mediated by additional machinery
to determine where the control flow would have gone in the original
program and redirecting it to the appropriate portion of the rewritten
program.

The approaches of REINS, SecondWrite and MULTIVERSE increasingly avoid
making decisions about code location and symbolization and thus offer
more guarantees to work for arbitrary binaries.  However, these
approaches also have disadvantages. They introduce overhead in the
rewritten binaries both in terms of speed and size.  Moreover, the
additional translation process for indirect jumps or calls is likely
to hinder later analyses on the disassembled code.  On
the other hand, our approach, although not guaranteed to work,
generates assembly code with symbolic references. This enables
performing advanced static analyses on the assembly code that can be
used to support more sophisticated rewriting techniques.
A binary can be rewritten multiple times without introducing a new
layer of indirection in every rewrite.



\subsection{Static Analysis Using Datalog}

Datalog has a long history of being used to specify and implement
static analyses. In 1995 Reps~\cite{Reps1995} presented an
approach to obtain demand driven dataflow analyses from the exhaustive
counterparts specified in Datalog using the magic sets
transformation. Much of the subsequent effort has been in scaling
Datalog implementations. In that vein, Whaley et
al.~\cite{Whaley:2004,Whaley} achieved significant pointer analysis
scalability improvements using an implementation based on binary
decision diagrams.  More recently, Datalog-based program analysis has
received new impetus with the development of Souffle~\cite{souffle},
a highly efficient Datalog engine.  The most prominent
application of Datalog to program analysis to date has been
Doop~\cite{Bravenboer,Smaragdakis,DatalogProgramAnalysis}, a context
sensitive pointer analysis for Java bytecode that scales to large
applications. Doop is currently one of the most comprehensive and
efficient pointer analysis for Java.

In the context of binary
analysis, we are only aware of the work of Brumley et
al.~\cite{brumley2006alias} which uses Datalog to specify an alias
analysis for assembly code.
Schwartz et al.~\cite{Schwartz:2018} present a binary analysis to
recover C++ classes from executables written in Prolog. Prolog, being
more expressive than Datalog, is typically evaluated starting from a
goal---in contrast to Datalog which can be evaluated bottom-up---and
using backtracking. Thus, in Prolog programs the order of the inference rules
is important and its evaluation is harder to parallelize.

Very recently, Grech et al.~\cite{gigahorse} have implemented a
decompiler, named Gigahorse, for Etherium virtual machine (EVM) byte
code using Datalog. Gigahorse shares some high level ideas with our
approach, i.e. the inference of high level information from low-level
code using Datalog. However, both the target and the inferred
information differ considerably. In EVM byte code, the main challenge
is to obtain a register based IR (EVM byte code is stack based),
resolve jump targets and identify function boundaries. On the other
hand, \ddisasm{} focuses on obtaining instruction boundaries and
symbolization information for x64 binaries.  Additionally, although
Gigahorse also implements heuristics using Datalog rules, it does not
use our approach of assigning weights to heuristics and aggregating
them to make final decisions.

\section{Preliminaries}
\subsection{Introduction to Datalog}

\begin{figure*}[t]
  \begin{tabular}{l|l}
    \begin{lstlisting}
instruction(A:$\address$,Size:$\num$,Prefix:$\str$,Opcode:$\str$,Op1:$\operand$,Op2:$\operand$,Op3:$\operand$,Op4:$\operand$)
invalid(A:$\address$)
op_regdirect(Op:$\operand$,Reg:$\register$)
op_immediate(Op:$\operand$,Immediate:$\num$)
op_indirect(Op:$\operand$,Reg1:$\register$,Reg2:$\register$,Reg3:$\register$,Mult:$\num$,Disp:$\num$,Size:$\num$)
    \end{lstlisting}
    &
    \begin{lstlisting}
data_byte(A:$\address$,Val:$\num$)
address_in_data(A:$\address$,Val:$\num$)
    \end{lstlisting}
    \end{tabular}
  \caption{Initial facts. Facts generated for executable sections on the left
  and facts generated for all sections on the right.}
  \label{fig:facts}
\end{figure*}

 A Datalog program is a collection of Datalog rules. A Datalog rule is
 a restricted kind of horn clause with the following format: $h
 \mathbin{:-} t_1, t_2, \dots, t_n$ where $h, t_1, t_2, \ldots, t_n$
 are predicates. Rules represent a logical entailment: $t_1\land t_2
 \land \ldots \land t_n \to h$.  Predicates in Datalog are limited to
 flat terms of the form $t(s_1, s_2, \ldots, s_n)$ where $s_1, s_2
 \ldots, s_n$ are variables, integers or strings.  Given a Datalog
 rule $h \mathbin{:-} t_1, t_2, \dots, t_n$, we say $h$ is the head of
 the rule and $t_1, t_2, \dots, t_n$ is its body.

Datalog rules are often recursive, and they can contain negated
predicates, represented as $! t$. However, negated predicates need to
be \emph{stratified}---there cannot be circular dependencies that
involve negated predicates e.g. $p(X) \mathbin{:-} ! q(X)$ and $q(A)
\mathbin{:-} ! p(A)$. This restriction guarantees that its semantics
are well defined. Additionally, all variables in a Datalog rule need
to be \emph{grounded}, i.e. they need to appear in at least one non-negated
predicate on the
rule's body.  Datalog also admits
disjunctive rules denoted with a semicolon e.g.  $h \mathbin{:-} t_1
\mathbin{;} t_2$ that are equivalent to several regular rules $h
\mathbin{:-} t_1$ and $h \mathbin{:-} t_2$.

The Datalog dialect that we adopt (Souffle's dialect) supports
additional constructs such as arithmetic operations, string operations
and aggregates. Aggregates compute operations over a complete set
of predicates such as summation, maximums or
minimums, and  we use them to integrate
the results of our heuristics.

A Datalog engine takes as input a set of facts, which are predicates
known to be true, and a
Datalog program (a set of rules). The engine generates new
predicates by repeatedly applying the inference rules until a fixpoint
is reached.  One of the appeals of Datalog is that it is fully
declarative. The result of a computation does not
depend on the order in which rules are considered or the order in
which predicates within a rule's body are evaluated. This makes it
easy to define multiple analyses that depend and collaborate with each
other.

In our case, the initial set of facts encodes all the information
present in the binary, the disassembly procedure (with all its
auxiliary analyses) is specified as a set of Datalog rules. The
results of the disassembly are the new set of predicates. These predicates are then used to
build an IR for binaries that can be
reassembled.

\subsection{Encoding Binaries in Datalog}

The first step in our analysis is to encode all the information
present in the binary into Datalog facts. We consider two basic
domains: strings, denoted as $\str$, and 64 bit machine numbers,
denoted as $\num$. We consider also the following sub-domains:
addresses $\address \subseteq \num$, register names
$\register\subseteq \str$ and operand identifiers
$\operand\subseteq\num$.  We adopt the convention of having Datalog
variables start with a capital letter and predicates with lower case.
We represent addresses in hexadecimal and all other numbers in
decimal. We only use the prefix |0x| for hexadecimal numbers if there
is ambiguity.

Fig.~\ref{fig:facts} declares the predicates used to represent the
initial set of raw instruction facts.  Predicate fields are annotated
with their type.  To generate these initial facts we apply a decoder
(Capstone~\cite{capstone}) to attempt to decode \emph{every} address
$x$ in the executable sections of a binary\footnote{This is
  different from linear disassembly which would try to decode an
  instruction at address $x+s$ after decoding an instruction of size
  $s$ at address $x$ (skipping the addresses in between).}.  If the
decoder succeeds, we generate an |instruction| fact with |A|$=x$. If
the decoder fails, the fact |invalid($x$)| is generated instead.  In each
|instruction| predicate, the field |Size| represents the size of the
instruction, |Prefix| is the instruction's prefix, and |Opcode| is the
instruction code.  Instruction operands are stored as independent
facts |op_regdirect|, |op_immediate| and |op_indirect|, whose first
field |Op| contains a unique identifier. This identifier is used to
match operands to their instructions. The fields |Op1| to |Op4| in
predicate |instruction| contain the operands' unique identifiers or
$0$ if the instruction does not have as many operands. We place source
operands first and the destination operand last. The predicate
|op_regdirect| contains a register name |Reg|, |op_immediate| contains
an immediate |Immediate| and |op_indirect| represents an indirect
operand of the form |Reg1:[Reg2+Reg3$\times$Mult+Disp]|. That is,
|Reg1| is the segment register, |Reg2| is the base register, |Reg3| is
the index register, |Mult| represents the multiplier, and |Disp|
represents the displacement. Finally, the field |Size| represents the
size of the data element being accessed in bytes.

\begin{figure}
  \lstset{language=[x64]Assembler}
  \begin{lstlisting}
416C35:   mov RBX, -624
416C3C:   nop
416C40:   mov RDI, QWORD PTR [RIP+0x25D239]
416C47:   mov RSI, QWORD PTR [RBX+0x45D328]
416C4E:   mov EDX, OFFSET 0x45CB23
416C53:   call 0x413050
416C58:   add RBX, 24
416C5C:   jne 0x416C40
\end{lstlisting}
\caption{Assembly (before symbolization) extracted from wget-1.19.1
  compiled with Clang 3.8 and optimization -O2. This code reads 8 byte
  data elements at address \lstinline{416C47} within the address range $[45D0B8,
    45D328]$ and spaced every 24 bytes.}
\label{fig:ex1}
\end{figure}

\begin{example}
  Consider the code in Fig.~\ref{fig:ex1}.
The  encoding of the instructions at addresses |416C47|
and |416C58| together with their respective operands
can be found below:

\noindent
\begin{lstlisting}
instruction(416C47,7,'','mov',14806,538,0,0)
op_indirect(14806,'NONE','RBX','NONE',1,45D328,8)
op_regdirect(538,'RSI')

instruction(416C58,4,'','add',188,519,0,0)
op_immediate(188,24)
op_regdirect(519,'RBX')
\end{lstlisting}

\noindent Note that the operand identifiers have no particular meaning. They are
assigned to operands sequentially as these are encountered during the
decoding.
\end{example}

In addition to decoding every possible instruction, we encode every section (both
data and executable sections) as follows.
For each address |A| in
a section,  a fact
|data_byte(A,Val)| is generated where |Val| is
the value of the byte at address |A|. We also generate the facts
|address_in_data(A,Addr)| for each
address |A| in a section such that the values of the
bytes from |A| to |A|$+7$ (8 bytes)\footnote{Our analysis considers x64
  architecture.} correspond to an address |Addr|
that falls in the address range of a section in the binary.
These facts will be our initial candidates for symbolization.
Executable sections are also encoded this way to support binaries
that interleave data with code.

Finally, additional facts are generated from the section, relocation,
and symbol tables of the executable as well as a special fact
|entry_point(A:$\address$)| with the entry point of the
executable. Note that for libraries, function symbol predicates
are generated for all exported functions and they will be considered
as entry points.

\section{Instruction Boundary Identification}
\label{sec:instructions}
The predicate |instruction| contains all the possible
instructions that might be in the executable. \ibi{} amounts to deciding which of these are real
instructions.  

Our \ibi{} is based on three
steps:
\begin{enumerate}[noitemsep,nolistsep]
\item A backward traversal starting from |invalid| addresses.
\item A forward traversal that combines elements of
   \emph{linear-sweep} and
   \emph{recursive-traversal}.
\item A conflict resolution phase to discard spurious blocks.
  \end{enumerate}

Both the backward and forward traversals use the auxiliary predicates
|may_fallthrough(From:$\address$,To:$\address$)| and
|must_fallthrough(From:$\address$,To:$\address$)| to represent
instructions at address |From| that may fall through or must fall
through to an address |To|.  Fig.~\ref{fig:aux-predicates} contains
the rules that define both predicates\footnote{Some of the rules have
  been slightly adapted for presentation purposes.}.  An instruction
at address |From| may fall through to the next one at address
|From+Size| as long as it is not a return, a halt, or an unconditional
jump instruction. Rule~\ref{drule:may_fallthrough} depends in turn on
other auxiliary predicates that abstract away specific aspects of
concrete assembler instructions e.g. |return_operation| is simply
defined as |return_operation('ret')| for x64.  The predicate
|must_fallthrough| restricts |may_fallthrough| further by discarding
instructions that might not continue to the next instruction
i.e. calls, jumps, or interrupt operations (we consider instructions
with a loop prefix as having a jump to themselves).

The traversals also depend on other predicates whose
definitions we omit:
|direct_jump(From:$\address$,To:$\address$)|,
|direct_call(From:$\address$,To:$\address$)|,
|pc_relative_jump(From:$\address$,To:$\address$)|, and
|pc_relative_call(From:$\address$,To:$\address$)| represent
instructions at address |From| that have a direct or RIP-relative
jump or call to an address |To|.

\begin{figure}
\begin{drule}{drule:may_fallthrough}
\begin{lstlisting}
may_fallthrough(From,To):-
    instruction(From,Size,_,OpCode,_,_,_,_),
    To=From+Size,
    !return_operation(OpCode),
    !unconditional_jump_operation(OpCode),
    !halt_operation(OpCode).
\end{lstlisting}
\end{drule}
\begin{drule}{drule:must_fallthrough}
\begin{lstlisting}		
must_fallthrough(From,To):-
    may_fallthrough(From,To),
    instruction(From,_,_,OpCode,_,_,_,_),
    !call_operation(OpCode),
    !interrupt_operation(OpCode),
    !jump_operation(OpCode),
    !instruction_has_loop_prefix(From).
\end{lstlisting}
\end{drule} 
\caption{Auxiliary Datalog predicates used for traversal.}
\label{fig:aux-predicates}
\end{figure}

\begin{example}
  Consider the code in Fig.~\ref{fig:ex1}. The |mov| instruction
  at address |416C4E| generates the predicates
  |must_fallthrough(416C4E,416C53)| and
  |may_fallthrough(416C4E,416C53)| whereas the call instruction
  only generates |may_fallthrough(416C53,416C58)|.  This is
  because the function at address |413050| (the target of the call)
  might not return.  The call instruction also generates the predicate
  |direct_call(416C53,413050)|. 
\end{example}

\subsection{Backward Traversal}
Our backward traversal simply expands the amount of |invalid|
predicates through the implication that any instruction
unconditionally leading to an invalid instruction must itself be
invalid.

\begin{drule}{drule:back-traversal}
\begin{lstlisting}
invalid(From):-
    (must_fallthrough(From,To) ;
        direct_jump(From,To) ;
        direct_call(From,To) ;
        pc_relative_jump(From,To) ;
        pc_relative_call(From,To)),
    (invalid(To) ;
         !instruction(To,_,_,_,_,_,_,_)).
\end{lstlisting}
\end{drule}
\begin{drule}{drule:possible-ea}
\begin{lstlisting}
possible_effective_address(A):-
    instruction(A,_,_,_,_,_,_,_), !invalid(A).
\end{lstlisting}
\end{drule}

Rule~\ref{drule:back-traversal} specifies that an instruction at
address |From| that jumps, calls or must fall through to an address
|To| that does not contain an potential instruction or to an address
|To| that contains an invalid instruction is also invalid. The
predicate |possible_effective_address(A:$\address$)| contains the
addresses of the remaining instructions not discarded by |invalid|
(Rule~\ref{drule:possible-ea}).

\subsection{Forward Traversal}
The forward traversal follows an approach that falls between the two
classical approaches \emph{linear-sweep} and
\emph{recursive-traversal}.
It traverses the code recursively but is much more
aggressive than typical traversals in terms of the targets that it
considers. Instead of starting the traversal only on the targets of
direct jumps or calls, every address that appears in one of the
operands of the already traversed code is considered a possible
target. For example, in Fig.~\ref{fig:ex1}, as soon as the analysis
traverses instruction {\lstset{language=[x64]Assembler}
  \lstinline{mov EDX, OFFSET 0x45CB23}}, it will consider the address
|45CB23| as a potential target that it needs to explore.
Additionally, potential addresses appearing in the data  (instances of
predicate |address_in_data|) are also considered potential targets.

The traversal is defined with two mutually recursive predicates:
|possible_target(A:$\address$)| specifies addresses where we start
traversing the code and
|code_in_block_candidate(A:$\address$,Block:$\address$)| takes care of
the traversing and assigning instructions to basic blocks.  A predicate
|code_in_block_candidate(A:$\address$,Block:$\address$)| denotes that
the instruction address |A| belongs to the candidate code block that
starts at address |Block|.

\begin{figure}
\begin{drule}{drule:block1}
\begin{lstlisting}
code_in_block_candidate(A,A):-
    possible_target(A),
    possible_effective_address(A).
\end{lstlisting}
\end{drule}
\begin{drule}{drule:block2}
\begin{lstlisting}
code_in_block_candidate(A,Block):-
    code_in_block_candidate(Aprev,Block),
    must_fallthrough(Aprev,A),
    !block_limit(A).
\end{lstlisting}
\end{drule}
\begin{drule}{drule:block3}
\begin{lstlisting}
code_in_block_candidate(A,A):-
    code_in_block_candidate(Aprev,Block),
    may_fallthrough(Aprev,A),
    (!must_fallthrough(Aprev,A) ;
        block_limit(A)),
    possible_effective_address(A).
\end{lstlisting}
\end{drule}
\begin{drule}{drule:target1}
\begin{lstlisting}
possible_target(A):-
    initial_target(A).
\end{lstlisting}
\end{drule}
\begin{drule}{drule:target2}
\begin{lstlisting}
possible_target(Dest):-
    code_in_block_candidate(Src,_),
    (may_have_symbolic_immediate(Src,Dest) ;
        pc_relative_jump(Src,Dest) ;
        pc_relative_call(Src,Dest)).
\end{lstlisting}
\end{drule}
\begin{drule}{drule:target3}
\begin{lstlisting}
possible_target(A):-
    after_block_end(_,A).
\end{lstlisting}
\end{drule}

\caption{Block forward traversal rules.}
\label{fig:forward_traversal}
\end{figure}

The definition of these predicates can be found in
Fig.~\ref{fig:forward_traversal}.  The traversal starts with the
|initial_target| (Rule~\ref{drule:target1}) that contains the
addresses of: entry points, any existing function symbols, landing pad
addresses (defined in the exception information sections), the start addresses of
 executable sections, and \emph{all} addresses in |address_in_data|. This
last component implies that all the targets of jump tables or function
pointers present in the data sections will be traversed.

However, not all jump tables are lists of absolute addresses
(captured by |address_in_data|).  Sometimes jump tables are stored as
differences between two symbols i.e. \lstinline{Symbol1-Symbol2}. In
these tables, the jump target |Symbol1| is computed by loading
|Symbol2| first and then adding the content of the jump table entry. We
found this pattern in PIC code and in position dependent code compiled
with ICC (see App.~\ref{sec:jump-tables}). An approximation of these jump tables is
detected with ad-hoc rules and their targets are included in
|initial_target|.

A possible target, marks the beginning of a new basic block
candidate (Rule~\ref{drule:block1}).  The candidate block is then extended as long as
the instructions are guaranteed to fall through and we do not reach a
|block_limit| (Rule~\ref{drule:block2}). The predicate |block_limit|
over-approximates |possible_target| (it is computed the same way
but without requiring the predicate |code_in_block_candidate| in Rule~\ref{drule:target2}).
Rule~\ref{drule:block3} starts a new block if the instruction is not
guaranteed to fall through or if there is a block limit. That is
where the previous block ends.
Any addresses or jump/call targets that appear in a block candidate
are considered new possible targets (Rule~\ref{drule:target2}).
|may_have_symbolic_immediate| includes direct jumps and calls but also
other immediates.  E.g. instruction
{\lstset{language=[x64]Assembler} |mov EDX, OFFSET 45CB23|} generates
|may_have_symbolic_immediate(416C4E,45CB23)|. Note that this is much
more aggressive than a typical recursive traversal that would only
consider the targets of jumps or calls.
Finally, Rule~\ref{drule:target3} adds a linear-sweep component to the
traversal.  |after_block_end(End:$\address$,A:$\address$)| contains
addresses |A| after blocks that end with an instruction that cannot
fall through at |End| (e.g. an unconditional jump or a return). This
predicate skips any padding (e.g., contiguous NOPs) that might be
found after the end of the previous block.

It is worth noting that in our Datalog specification we do not have to
worry about many issues that would be important in lower level
implementations of equivalent binary traversals. For instance, we do not
need to keep track of which instructions and blocks have already been
traversed nor do we specify the order in which different paths are
explored.

\subsection{Solving Block Conflicts}
\label{sec:block-conflicts}
Once the second traversal is over, we have a set of candidate blocks,
each one with a set of instructions (encoded in the predicate
|code_in_block_candidate|).  These blocks represent our best effort to
obtain an over-approximation of the basic blocks in the original
program. In principle, it is possible to miss code blocks. However,
such code block would have to be reachable only through a computed
jump/call \emph{and} be preceded by data that derails the linear-sweep
component of the traversal (Rule~\ref{drule:target3}). We have not
found any instance of this situation. We remark that if the address of
a block appears anywhere in the code or in the data, it will be
considered.  For instance, ICC
puts some jump tables in executable sections. By detecting these
jump tables, we consider their jump targets (which are typically the blocks
after the jump table) as possible targets in our traversal.

The next step in our \ibi{} is to decide
which candidate blocks are real. For that, we detect the blocks that
overlap with each other or with a potential data segment (e.g. a jump
table in the executable section). Overlapping blocks are extremely uncommon
in compiled code. The situations in which they appear tend to respond to very
specific patterns such as a block starting with or without a
|lock| prefix \cite{meng2016binary}. We recognize those patterns
with ad-hoc rules and consider that the remaining blocks should not
overlap. Thus, if two blocks overlap, we assume one of them is
spurious and needs to be discarded.  This assumption could be relaxed
if we wanted to disassemble malware but it is generally useful for
 compiled binaries.

 We decide which blocks to discard using heuristics. Each
 heuristic is implemented as a Datalog rule that produces a predicate
 of the form
 \lstinline{block_points(Block:$\address$,Src:$\address$,Points:$\num$,Why:$\str$)}.
 Such a predicate
assigns |Points| points to the block starting at address
|Block|. The field |Src| is an optional reference to another block that is
the cause of the points or zero for heuristics that are not based on other blocks.
The field |Why| is a string that describes the
heuristic for debugging purposes and to distinguish the predicate
from others generated from different heuristics.

We compute the total number of points for each block using Souffle's
aggregates\cite{souffle}.  Then, given two overlapping blocks, we
discard the one with the least points. In case of a tie,
we keep the first block and emit a warning. We also discard blocks if their
total points is below a threshold. This is useful for blocks
whose heuristics indicate overlap with data elements.

Our heuristics are mainly based on how blocks are interconnected,
how they fit together spatially, and whether
they are referenced by potential pointers or overlap with jump tables.
Some of the heuristics used are described below ($+$ for positive points and $-$ for
negative points):
\begin{itemize}[nolistsep,noitemsep]
\pro The block is called, jumped to, or there is
  a fallthrough from a non-overlapping block.
\pro The block's initial address appears somewhere
  in the code or data sections. If the appearance is at an aligned address,
  it receives more points.
\pro The block calls/jumps other non-overlapping blocks.
\con A potential jump table
  overlaps with the block.
\end{itemize}

All memory not covered by a block is considered data.

\section{Auxiliary Analyses}
\label{sec:analysis}
The next step in our disassembly procedure is symbolization. However,
we first perform several static analyses to infer how data is accessed
and used, and thus deduce its layout.







\subsection{Register Def-Use Analysis}
\label{sec:def-use}
First, we compute register definition-uses chains. The analysis
produces predicates of the form:\\
\begin{tabular}{c}
\begin{lstlisting}[framesep=3pt]
def_used(Adef:$\address$,Reg:$\register$,Aused:$\address$,Index:$\num$)
\end{lstlisting}
\end{tabular}\\
The register |Reg| is defined at address |Adef| and
used at address |Aused| in the operand with index |Index|.

The analysis first infers definitions
|def(Adef:$\address$,Reg:$\register$)| and uses
|use(Aused:$\address$,Reg:$\register$,Index:$\num$)|. Then, it
propagates definitions through the code and matches them to uses. The
analysis is intra-procedural in that it does not traverse
calls but only direct jumps. This makes the analysis incomplete but
improves scalability. During the propagation of definitions, the
analysis assumes that certain registers keep their values through
calls following Linux x64 calling convention~\cite{abi}.

\begin{example}
  \label{ex:defuse}
  Consider the code fragment in Fig.~\ref{fig:ex1}. The Def-Use analysis
  produces the following predicates:\\
  \begin{tabular}{c}
  \begin{lstlisting}
def_used(416C35,'RBX',416C47,1)
def_used(416C35,'RBX',416C58,2)
def_used(416C58,'RBX',416C58,2)
def_used(416C58,'RBX',416C47,1)
  \end{lstlisting}
  \end{tabular}
\end{example}
One important detail is that the analysis considers the 32 bits and 64
bits registers as one given that the x64 architecture zeroes the upper
part of 64 bits registers whenever the corresponding 32 bits register
is written. That means that for instruction
{\lstset{language=[x64]Assembler}|mov EDX, OFFSET 0x45CB23|} at
address |416C4E|, the analysis generates a definition |def(416C4E,RDX)|.

Once we have def-use chains, we want to know which register
definitions are potentially used to compute addresses
to access memory. For that purpose, the disassembler computes
a new predicate:\\
\begin{tabular}{c}
\begin{lstlisting}[framesep=3pt]
def_used_for_address(Adef:$\address$,Reg:$\register$)
\end{lstlisting}
\end{tabular}\\
that denotes that the register |Reg| defined at address |Adef| might
be used to compute a memory access.  This predicate is computed by
traversing def-use chains backwards starting from instructions
that access memory.  This traversal is transitive, if a register $R$
is used in an instruction that defines another register $R'$ and that
register is used to compute an address, then $R$ is
also used to compute an address. This is captured
in the following Datalog rule:

\noindent
\begin{drule}{drule:used-for-address}
\begin{lstlisting}
def_used_for_address(Adef,Reg):-
    def_used_for_address(Aused,_),
    def_used(Adef,Reg,Aused,_).
\end{lstlisting}
\end{drule}

\subsection{Register Value Analysis}
\label{sec:value-analysis}
In contrast to instructions that refer to code, where direct
references (direct jumps or calls) predominate, memory accesses are
usually computed. Rather than accessing a fixed address, instructions
typically access addresses computed with a combination of register
values and constants. This address computation is often done over
several instructions. Such is the case in the example code in Fig.~\ref{fig:ex1}.

In order to approximate this behavior, we developed an analysis that
computes the value held in a register at an address.  There are many
ways of approximating register values ranging from simple
constant propagation to complex abstract domains that take memory
locations into account e.g.~\cite{vsa}. Generally, the more complex the
analysis domain, the more expensive it is.  Therefore,
we have chosen a minimal representation that captures the kind of
register values that are typically used for accessing memory. Our
value analysis representation is based on the idea that typical memory
accesses follow a particular pattern where the memory address that is
accessed is computed using a base address, plus an index multiplied by
a multiplier.  Consequently, the value analysis produces predicates of
the form:
\begin{tabular}{c}
  \begin{lstlisting}[framesep=3pt]
reg_val(A:$\address$,Reg:$\register$,A2:$\address$,Reg2:$\register$,Mult:$\num$,Disp:$\num$)
\end{lstlisting}
\end{tabular}
  
\noindent which represents that the value of a register |Reg| at
address |A| is equal to the value of another register |Reg2| at
address |A2| multiplied by |Mult| plus an
displacement |Disp| (or offset).

The analysis proceeds in two phases. The first phase produces
predicates of the form |reg_val_edge| which share the signature
with |reg_val|.  We generate one |reg_val_edge| per
instruction and def-use predicate for the instructions whose behavior
can be modeled in this domain and are used to compute an address (|def_used_for_address|).
For example, Rule~\ref{drule:value-ex} below generates |reg_val_edge| predicates for
|add| instructions that add a constant to a register:

\noindent
\begin{drule}{drule:value-ex}
\begin{lstlisting}
reg_val_edge(A,Reg,Aprev,Reg,1,Imm):-
    def_used_for_address(Aprev,Reg),
    def_used(Aprev,Reg,A,_),
    instruction(A,_,_,'add',Op1,Op2,0,0),
    op_immediate(Op1,Imm),
    op_regdirect(Op2,Reg).
\end{lstlisting}
\end{drule}

\begin{example}
  \label{ex:reg_val_edge}
  Continuing with Example~\ref{ex:defuse}, the predicates |reg_val_edge| 
  generated for the code in Fig.~\ref{fig:ex1} are:\\
   \begin{tabular}{@{}c@{}}
  \begin{lstlisting}[mathescape=true]
$\pcount{val1}$ val_reg_edge(416C35,'RBX',416C35,'NONE',0,-624)
$\pcount{val2}$ val_reg_edge(416C58,'RBX',416C35,'RBX',1,24)
$\pcount{val3}$ val_reg_edge(416C58,'RBX',416C58,'RBX',1,24)
  \end{lstlisting}
  \end{tabular}
  Predicate $P\ref{val1}$ captures that |RBX| has a constant value after executing
  the instruction in address |416C35| (note that the multiplier is $0$ and the
  register has a special value |'NONE'|). Predicate $P\ref{val2}$, generated
  from Rule~\ref{drule:value-ex}, specifies
  that the value of |RBX| defined at address |416C58| corresponds to
  the value of |RBX| defined at |416C35| plus $24$.
  Finally,  $P\ref{val3}$ denotes that the value of |RBX|
  at |416C58| can be the result of incrementing the value of |RBX|
  defined at the same address by $24$.
\end{example}
The set of predicates |reg_val_edge| can be seen as directed
relational graph. The nodes in the graph are pairs of address and register
|(A, Reg)| and the edges express relations between their values i.e.
they are labeled with a multiplier and offset.

Once this graph is computed, we perform a propagation phase
akin to a transitive closure. This propagation phase chains together
|reg_val_edge| predicates.  The chaining starts from the leafs of
the graph (nodes with no incoming edges). Leafs in the
|reg_val_edge| graph can be instructions that load a constant into a
register such as {\lstset{language=[x64]Assembler}|mov RBX, -624|}
in Fig.~\ref{fig:ex1} or instructions
where a register is assigned the result of an operation not supported by the domain.
For example, loading a value from memory {\lstset{language=[x64]Assembler}
|mov RDI, [RIP+0x25D239]|}
in Fig.~\ref{fig:ex1}. In that case, the generated predicate would be the
tautological predicate |reg_val(416C40,RBX,416C40,RBX,1,0)|.

In order to ensure termination and for efficiency reasons we
limit the number of propagation steps by a constant |step_limit| with
an additional field |S:$\num$| in the |reg_val| predicates.  The
main rule for combining |reg_val_edge| predicates is the
following:\\
\begin{drule}{drule:valpropagation}
\begin{lstlisting}
reg_val(A1,R1,A3,R3,M1*M2,(D2*M1)+D1,S+1):-
    reg_val(A2,R2,A3,R3,M2,D2,S),
    reg_val_edge(A1,R1,A2,R2,M1,D1), A1 != A2,
    step_limit(Limit), S+1 < Limit.
\end{lstlisting}
\end{drule}
This rule chains edges linearly by combining their multipliers
and displacements. It keeps track of
operations that involve one source register and one destination
register. However, we also want to detect
situations where multiple edges converge into one instruction.
Specifically, we want to detect loops
and operations that involve multiple registers.

\textbf{Detecting Simple Loops.}
The following rule (Rule~\ref{drule:loop}) detects situations where a
register |R| is initialized to a constant |D1|, then
incremented/decremented in a loop by a constant |D2|.\\
\noindent\begin{drule}{drule:loop}
  \begin{lstlisting}
reg_val(A,Reg,A2,'Unknown',D2,D1,S+1):-
    reg_val(A,R,A2,'NONE',0,D1,S),
    reg_val_edge(A,R,A,R,0,D2),
    step_limit(Limit), S+1 < Limit.
  \end{lstlisting}
\end{drule}

This pattern
can be interpreted as |D1| being the base for a memory address and
|D2| being the multiplier used to access different elements of a data
structure. Our new multiplier |D2| does not actually multiply any real
register, so we set the register field to a special value |'Unknown'|.

\begin{example}
  Consider the propagation of the predicates in
  Example~\ref{ex:reg_val_edge}. The generated predicates are:\\
  \begin{tabular}{c}
    \begin{lstlisting}[mathescape=true]
$\pcount{val4}$ val_reg(416C35,'RBX',416C35,'NONE',0,-624)
$\pcount{val5}$ val_reg(416C58,'RBX',416C35,'NONE',0,-600)
$\pcount{val6}$ val_reg(416C58,'RBX',416C35,'Unknown',24,-600)
    \end{lstlisting}
  \end{tabular}
  First,  predicate $P$\ref{val4} is
  generated from  $P$\ref{val1} which is a leaf.  Then, $P$\ref{val4} is combined
  with  $P$\ref{val2} using
  Rule~\ref{drule:valpropagation} into predicate
  $P$\ref{val5}.  Finally, Rule~\ref{drule:loop} is
  applied to $P$\ref{val5} and  $P$\ref{val3} to generate
  $P$\ref{val6} which denotes that the
  register |RBX| takes values that start at $-600$ and are incremented
  in steps of $24$ bytes.
\end{example}

\textbf{Multiple Register Operations.}
In general, operations over two source registers cannot be expressed with
|reg_val| predicates. However, if one of the registers has a constant value
or both registers can be expressed in terms of a third common register (a diamond
pattern), we can propagate their value.

\begin{example}
  \label{ex:diamond}
  The following assembly code contains a simple diamond pattern:\\
\begin{tabular}{l}
  { \lstset{language=[x64]Assembler}
  \begin{lstlisting}
0: mov RBX, [RCX]
1: mov RAX, RBX
2: add RAX, RAX
3: add RAX, RBX
  \end{lstlisting}}
 \end{tabular}\\
  The last instruction adds the registers |RAX| and |RBX|. However,
  the value of |RAX| is two times the value of |RBX|.
  This is reflected in the predicates |reg_val(2,RAX,0,RBX,2,0)|
  and |reg_val(0,RBX,0,RBX,1,0)|.
  Therefore, we can generate a predicate
  |reg_val(3,RAX,0,RBX,3,0)|.
\end{example}

Note that the register value analysis intends to capture some of the
relations between register values but it makes no attempt capture all
of them. The goal of this analysis is not to obtain a sound
over-approximation of the register values but to provide as much
information as possible about how memory is accessed. The analysis is
also not strictly an under-approximation as it is based on def-use
chains which are over-approximating.

\subsection{Data Access Pattern Analysis}
\label{sec:data-access}
The data access pattern (\dap{}) analysis takes the results of the register
value analysis and the results of the def-use analysis
to infer the register values at each of the
data accesses and thus compute which addresses are accessed and  which
pattern is used to access them.
The \dap{} analysis generates predicates of the form:\\
\begin{tabular}{c}
\begin{lstlisting}[framesep=3pt]
data_access_pattern(A:$\address$,Size:$\num$,Mult:$\num$,From:$\address$)
\end{lstlisting}
\end{tabular}
which specifies that address |A| is accessed from an instruction
at address |From| and  |Size| bytes are read or written. Moreover,
the access uses a multiplier |Mult|.

\begin{example}
  The code in Fig.~\ref{fig:ex1} generates several \dap{}s:\\
    \begin{tabular}{c}
    \begin{lstlisting}
$\pcount{dap1}$ data_access_pattern(673E80,8,0,416C40)
$\pcount{dap2}$ data_access_pattern(45D0B8,8,0,416C47)
$\pcount{dap3}$ data_access_pattern(45D0D0,8,24,416C47)
  \end{lstlisting}
  \end{tabular}\\
  The instruction at address |416C40| produces $P$\ref{dap1}
  which represents an access to a fixed address that reads 8 bytes.
  Conversely, the instruction at address |416C47| yields
  two predicates: $P$\ref{dap2} and $P$\ref{dap3}.
  This is because register |RBX| can have multiple values at address
  |416C47|. If there are multiple \dap{}s to the same
  address, we choose the one with the highest multiplier.
\end{example}

These \dap{}s provide very sparse information, but if an
address $x$ is accessed with a multiplier $m$, it is likely that
$x+m$, $x+2m$, etc., are also accessed the same way. Thus, we extend
\dap{}s based on their multiplier.  The analysis produces
a predicate |propagated_data_access| with the same format as
|data_access_pattern|.  Our auxiliary analyses provide no information
on what is the upper limit of an index in a data access.  Thus, we
simply propagate a \dap{} until it reaches the next \dap{}
that coincides on the same address or that has a
different multiplier.
The idea behind this criterion is that the next data structure in the
data section is probably accessed from somewhere in the code. So
rather than trying to determine the size of the data structure being
accessed, we assume that such data structure ends where the next one
starts.  These propagated \dap{}s will inform our
symbolization heuristics.

\begin{example}
  In our running example (Fig.~\ref{fig:ex1}) the \dap{}
  |data_access_pattern(45D0D0,8,24,416C40)| is propagated from
  address |45D0D0| up to address |45D310| in $24$ byte intervals. The generated predicates
  are:\\
  \begin{tabular}{l}
  \begin{lstlisting}
    propagated_data_access(45D0D0,8,24,416C40)
    propagated_data_access(45D0E8,8,24,416C40)
       $\cdots$                                 $\cdots$
    propagated_data_access(45D310,8,24,416C40)
  \end{lstlisting}
  \end{tabular}
  The \dap{} is not propagated to the next address |45D328|
  because that address contains another \dap{} generated
  at a different part of the code.
\end{example}

\subsection{Discussion}
\label{sec:discussion}

There are two important aspects that set our register value analysis
and \dap{} analysis apart from previous approaches like
\Ramblr{}~\cite{ramblr}.

First, the register value analysis is relational---it represents the
value of one register at some location in terms of the value of
another register at \emph{another} location---in contrast to
traditional value set analyses (VSA)~\cite{vsa}.  This is also
different from the affine-relations analysis~\cite{affine-relations}
used in VSA analyses which computes relations between register values
at the \emph{same} location.  A |reg_val| predicate between two
registers also implies a data dependency i.e. a register is defined in
terms of the other.

As a consequence, register value analysis can provide useful
information (for our use-case) in many cases where obtaining a
concrete value for a register would be challenging.
Consider the code
in Example~\ref{ex:diamond}. Our analysis concludes that at address
$3$ |RAX| is $3$ times the value of |RBX| at address $0$ regardless of
what that value might be.  In contrast, a traditional VSA analysis will only
provide useful information for the value of |RAX| as long as it can
precisely approximate the value of |RCX| and the values of all the
possible memory locations pointed by |RCX|.  If any of those locations
has an imprecise abstract value e.g. $\top$, so will |RAX|.

\begin{example}
Let us consider a continuation of Example~\ref{ex:diamond}:\\
\begin{tabular}{l}
  {
  \lstset{language=[x64]Assembler}
\begin{lstlisting}
4: mov R8, QWORD PTR [RAX*8+0x1000]
5: mov R9, WORD PTR [RAX*8+0x1008]
6: mov R10, BYTE PTR [RAX*8+0x1010]
\end{lstlisting}
  }
\end{tabular}\\
There will be \dap{}s for addresses |0x1000|, |0x1008| and
|0x1010| with sizes $8$, $2$, and $1$ and a multiplier of $24$ each.  This
information, though unsound in the general case (we are assuming |RAX|
can take the value $0$), is useful in practice.
\end{example}

These \dap{}s are the second distinguishing aspect of our
analyses. \Ramblr{} recognizes primitives and arrays of
primitives. However, these \dap{}s indicate that address |0x1000|
likely contains a |struct| with (at least) three fields of different
sizes. Moreover, thanks to the multiplier and the
|propagated_access_pattern| predicate we can conclude that address
|0x1000| holds in fact an array of structs where the first field (at
addresses |0x1000|, |0x1018|, |0x1030|$\ldots$) has size $8$ and might contain a
pointer whereas the second and third fields (at addresses |0x1008|, |0x1020|,
|0x1038|$\ldots$ and |0x1010|, |0x1028|, |0x1040|$\dots$ respectively)
have size $2$ and $1$ and thus are unlikely to hold a pointer.

\section{Symbolization}
The next step to obtain assembleable code is to perform
symbolization.  It
consists of deciding for each constant in the code or  in
the data  whether it is a literal or a symbol.  A first
approximation can be achieved by considering as symbols all
numbers that fall within the range of the address space. However, as
reported by Wang et al.~\cite{ramblr}, this leads to both false
positives and false negatives. Next, we explain our approach to reduce
the presence of false positives and negatives.

\subsection{False Positives: Value Collisions}
False positives are due to value collisions, literals that happen to
coincide with range of possible addresses.  In order to reduce the
false positive rate, we require additional
evidence in order to classify a number as a symbol.

\subsubsection{Numbers in Data}
For numbers in data, similarly to the approach used for
blocks, we start by defining a set of ``data object'' candidates.
Each candidate has an address, a type, and a size.
We define data object candidates for the following types:
\begin{description}[nolistsep,noitemsep]
\item[Symbol] Whenever
  the number falls into the right range (|address_in_data|).
\item[String] A sequence of printable characters ended in $0$.
\item[Symbol-Symbol] We detect jump tables using ad-hoc rules based on def-use chains,  register
values, and the \dap{}s computed in Sec.~\ref{sec:analysis} (see App.~\ref{sec:jump-tables}).
\item[Other] An address is accessed with a
  different size than the pointer size (8 bytes in x64 architecture)
  using the predicate |propagated_data_access| computed in
  Sec.~\ref{sec:data-access}.
\end{description}

We assign points to each of the candidates using heuristics based on
the analyses results and detect if they are overlapping. If they are,
we discard the candidate with fewer points.  This process is analogous
to how conflicts are resolved among basic blocks in
Sec.~\ref{sec:block-conflicts}. Note that detecting objects of type
``String'' and ``Other'' helps to discard false positives (i.e. symbol
candidates) that overlap with them.  As with blocks, we discard
candidates if their total points fall below a threshold.

The main heuristics for data objects are ($+$ positive
points and $-$ for negative points):
\begin{itemize}[noitemsep,nolistsep]
  \pro \textbf{Pointer to instruction beginning:} A symbol candidate
  points to the  beginning of an
  instruction.  This heuristic relies on the results of the already
  computed \ibi{}.
\pro \textbf{Data access match:} The data object candidate is accessed from
  the code with the right size.  This heuristic
  checks the existence of a |propagated_data_access| that matches
  the data object candidate's address and size.
\pro \textbf{Symbol arrays:} There are several (at least 3)  contiguous or evenly
spaced symbol candidates. This indicates that they belong to
the same data structure. Also, it is less
likely to have several consecutive value collisions.
\pro \textbf{Pointed by symbol array:} Multiple candidates of the same
type pointed by a single symbol array.
\pro \textbf{Aligned symbols:} A symbol
  candidate is located at an address with $8$ bytes alignment.
  \pro \textbf{Strings:} A string candidate receives some points
  by default. If the string is longer that $5$ bytes, it receives more points.
\con \textbf{Access conflict:} There is some data access in the middle
of a symbol candidate.
\con \textbf{Pointer to special section:} A symbol candidate points to
a location inside a special section such as |.eh_frame|.
\end{itemize}

\subsubsection{Numbers in Code}
We follow the same approach to disambiguate numbers in
instruction operands. However, only the first and the last heuristics of the ones listed above,
``Pointer to instruction beginning''
and ``Pointer to special section,'' are applicable to numbers in
code. We distinguish two cases: numbers that represent immediate
operands and numbers that represent a displacement in an indirect operand.
After taking these two heuristics into account,
we have not found false positives in displacements.  For
immediate operands we consider the following additional heuristics:
\begin{itemize}[noitemsep,nolistsep]
  \pro \textbf{Used for address:} The immediate is stored in a
  register used to compute an address (detected using predicate
  |def_used_for_address| from Sec.~\ref{sec:analysis}).

\con \textbf{Uncommon pointer operation:} The immediate
  or the register where it is loaded is used in an operation
  uncommon for pointers such as |MUL| or |XOR|.

\con \textbf{Compared to non-address:} The immediate is compared or moved
to a register that in turn is compared to another immediate that cannot
be an address.
\end{itemize}
These heuristics are tailored to the inference of how the immediate is
used, and they rely on def-use chains and the results of the register
value analysis.

\subsection{False Negatives: Symbol+Constant}
\label{sec:false-negatives}
False negatives can occur in situations where the original code
contains an expression of the form |symbol+constant|. In such cases,
the binary under analysis contains the result of computing that
expression.

There is no general procedure to recover the original expression
in the code as that information is simply not present in the binary.
Having a new symbol pointing to the
result of the |symbol+constant| expression instead of the original expression
is not a problem for rewrites which leave the data
sections unmodified (even if the sections are moved) or rewrites that only add data to the beginning or the end of data
sections. However, sometimes the resulting address of a
 |symbol+constant| expression falls outside the data section ranges
or falls into the wrong data section. In such cases, a naive symbolization
approach can result in false negatives.

We detect and correct these cases by detecting common
patterns where compilers generate |symbol+constant| using the
results of our def-use analysis and the register value analysis.  We
distinguish two cases: displacements in an indirect operands and
immediate operands.

\subsubsection{Displacements in Indirect Operands}
For displacements in indirect operands, we know that the address that
results from the indirect operand should be valid.  Consider a generic
data access |[R1+R2$\times$M+D]| where |R1| and |R2| are registers, |M| is
the multiplier and |D| the displacement.  The displacement |D| might
not fall onto a data section, but the expression |R1+R2$\times$M+D|
should.

Typically, in a data access as the one above, one of the addends
represents a valid base address that points to the beginning of a data
structure and the rest of the addends represent an offset into the
data structure.  In our generic access, |D| might be the base address,
in which case it should be symbolic, or the base address might be in
one of the registers, in which case |D| should not be symbolic.

We detect cases in which |D| should be symbolic even if it does not
fall in the range of a data section. For example if the data access is
of the form |[R2$\times$M+D]| with |M|$\,>1$, it is likely that |D|
represents the base address and should be symbolic.  We can detect
less obvious cases with the help of the register value analysis (see
Sec.~\ref{sec:value-analysis}). If we have a data access of the
form |[R1+D]| but the value of |R1| can be expressed as the value of
some other register |Ro| multiplied by a multiplier |M|$\,>1$ (there is a
predicate of the form |reg_val(_,R1,_,Ro,M,0)|) , then |D| is also
likely to be the base address and thus symbolic.  On the other hand,
if |R1| has a value that is a valid data address
(there is a predicate |reg_val(_,R1,_,'NONE',0,A)| where |A| falls
in a data section), then |D| is probably not a base address.

Knowing that a displacement should be symbolic is not enough, we need
to infer the right data section to which the symbolic expression
should refer. If the data access generates a \dap{}, we
use the destination address of the \dap{} as a reference for creating
the symbolic expression. Otherwise, we choose the closest boundary of a
data section as a reference.

\subsubsection{Immediate Operands}
Having a symbolic immediate that falls outside the data sections
is uncommon. The main pattern that we have identified is
when the immediate is used as an initial value for a loop counter
or as a loop bound to which the counter is compared.

\begin{example}
  \label{example:immediateConstant}
  Consider the following code fragment taken from the program
  conflict-6.0 compiled with GCC 5.5 and optimization -O1.
  It presents an immediate of the form \lstinline{symbol+constant}
  landing in a different section.\\
\begin{tabular}{c}
\lstset{language=[x64]Assembler}  
\begin{lstlisting}
40109D:   mov EBX, 402D40
4010A2:   mov EBP, 402DE8
4010A7:   mov RCX,QWORD PTR [RBX]
 ...       ...
4010C5:   add RBX,8
4010C9:   cmp RBX,RBP
4010CC:   jne 4010A7
\end{lstlisting}
\end{tabular}\\
  The number |402DE8| loaded at
  |4010A2| represents a loop bound and it is used in instruction
  |4010C9| to check if the end of the data
  structure has been reached.  Address |402d40| is in
  section {\lstinline{.rodata}} but
  address |402DE8| is in section
  \lstinline{.eh_frame_hdr}.
\end{example}

We detect this and similar patterns by combining the information
of the def-use analysis and the value analysis.  We note that in these
situations, the address that falls outside the section or on a
different section and the address range of the correct section are within
the distance of one multiplier.  That is, let $x$ be a candidate
address that might represent the result of a |symbol+constant|
expression, and let $[s_i,s_f)$ be the address range of the
original symbol's section.  Then $x\in [s_i-M, s_f+M]$ where $M$ is
the increment of the loop counter.  Therefore, our detection
mechanism generates an extended section range as above for every register
that we identify as loop counter.
Then, it checks if there is some immediate compared to the
loop counter that falls within this extended range. If that happens,
the immediate is rewritten using the base of the loop counter as a
symbol.

    


\begin{example}
  Example~\ref{example:immediateConstant} continued.  The
  register value analysis detects that |RBX| is a loop counter
  with a base address of |402D40| and a step size of |8|.
  Thus, we consider an extension of section |.rodata| to the
  range |[402718, 402DF0]| (the original address range is |[402720, 402DE8)|).
  Finally, using def-use chains we detect that the loop counter is compared
  to the immediate |402DE8| which falls within the extended section range.
  Consequently, we generate the following statement:\\
    {
    \lstset{language=[x64]Assembler}
    \begin{tabular*}{\linewidth}{c}
      |4010A2: mov EBP,OFFSET .L_402D40+168|
    \end{tabular*}\\
  where |.L_402D40|} is a new symbol pointing to address
   |402D40|.
\end{example}

\begin{table*}[t]
\begin{center}
\begin{tabular}{ll|ll|ll|ll|ll}
Program & Size & Program & Size & Program & Size & Program & Size& Program & Size\\ \hline
bar-1.11.0   & 91 &
bison-2.1    & 359 &
bool-0.2     & 48  &
conflict-6.0 & 28  & 
doschk-1.1   & 18  \\ 
ed-0.9       & 63  &
enscript-1.6.1 & 253 &
flex-2.5.4   & 196 & 
gawk-3.1.5   & 485 &
gperf-3.0.3  & 409 \\
grep-2.5.4   & 181 &
gzip-1.2.4   & 81 &
lighttpd-1.4.18 & 255 &
m4-1.4.4     & 154 &
make-3.80    & 202 \\
marst-2.4    & 104 & 
patch-2.6.1  & 155 &
re2c-0.13.5  & 2554 & 
rsync-3.0.7  & 1685 &
sed-4.2     & 201 \\
tar-1.29     & 547 &
tnef-1.4.7   & 74 &
units-1.85   & 65 &
wget-1.19.1  & 620 &
yasm-1.2.0   & 899\\
\end{tabular}
\end{center}
\caption{Real world example benchmarks. Each program is
    annotated with its size in KB when compiled with GCC 7.1.0 and
    optimization flag -O0.}
\label{table:real_world}
\end{table*}

\begin{table*}[hbt]
  \centering
  \begin{tabular}{lrr|  rrrr|  rrrr}
   
    \multirow{2}{*}{Benchmark}  & \multirow{2}{*}{Binaries} &
    \multirow{2}{*}{Refs} &
     \multicolumn{4}{c}{\ddisasm{}} & \multicolumn{4}{|c}{\Ramblr{}}\\
    &&& \multicolumn{1}{l}{FP} & FN & WS & Broken  & FP & FN  & Broken & Broken w/o ICC\\
     \hline
    Real world    & 1050    & 5957016   & 0  & 20  &  50  & 6   & 50258 & 62060 & 408 & 273  \\
    Coreutils     & 3710   & 4279339   & 3  & 0   & 0    & 3   & 8246 & 140774 & 752 & 323  \\
    CGC           & 2898    & 7220451   & 0  & 17  &  2   & 12   & 10892 & 43683 & 391 & 31 \\
  \end{tabular}
  \vspace{-5pt}
  \caption{Symbolization evaluation of \ddisasm{} and \Ramblr{}.
      ``Refs'' represents the total number of references in these binaries; ``FP'' and
      ``FN'' list the number of false positives and false negatives
      respectively for each tool; ``WS'' lists the number of
      references pointing to the wrong section (only shown for
      \ddisasm{}); ``Broken'' lists the number of binaries
      that are broken (have at least one ``FP,'' ``FN'' or
      ``WS''). ``Broken w/o ICC'' lists  broken binaries
  without counting the ones compiled with ICC.}
  \label{table:symbolization}
\end{table*}

\section{Experimental Evaluation}
We implemented our disassembly technique in a tool called \ddisasm{}.
\ddisasm{} takes a binary and produces an IR
called \ifanonymous \gtirb{}. \else
GrammaTech Intermediate Representation for Binaries (\gtirb{})~\cite{gtirb}. \fi
This representation can be
printed to assembly code that can be directly reassembled.  Currently
\ddisasm{} only supports x64 Linux ELF binaries but we plan to
extend it to support other architectures and binary formats.
\ddisasm{} is predominantly implemented in Datalog ($4336$ non-empty LOC) which is compiled
into highly efficient parallel C++ code using Souffle~\cite{souffle}.

\textbf{Benchmarks.}  We performed several experiments against a
variety of benchmarks, compilers, and optimization flags.  We selected
$3$ benchmarks. The first one is \texttt{Coreutils 8.25} which is
composed of 106 binaries and has been used in the experimental
evaluations of \Ramblr~\cite{ramblr} and Uroboros~\cite{uroboros}.
Programs in Coreutils are known to share a lot of
code~\cite{nucleous}, so it is important to also consider other
benchmarks.
The second benchmark is a subset of the programs from the DARPA Cyber
Grand Challenge (CGC). We adopt a modified version of these binaries
that can be compiled for Linux systems in x64~\ifanonymous\cite{cgc-grammateech-anonymous}\else\cite{cgc-grammateech}\fi.
We exclude programs that fail to compile or fail all their
tests. That leaves $69$ CGC
programs.
Finally, the third benchmark is a collection of $25$ real world open
source applications whose binary size ranges from $28$~KB to
$2.5$~MB.
Table~\ref{table:real_world} contains the names,
version, and sizes (in KB) of the applications in the
real world benchmark.
Some of the original binaries in all benchmarks fail some
tests. We take the
results of the original binary as a baseline which
rewritten binaries must match exactly---including failures.

\textbf{Compilation Settings.}  For each of those programs we compile
the binaries with $7$ compilers: GCC 5.5.0, GCC 7.1.0, GCC 9.2.1, Clang
3.8.0, Clang 6.0, Clang 9.0.1, and ICC 19.0.5.  For each compiler we use the
following $6$ compiler flags: -O0, -O1, -O2, -O3, -Os, and -Ofast.  All
programs are compiled as \emph{position dependent code}\footnote{This
  is harder to disassemble than position {\it independent} code (PIC),
  which is though to be easier because it contains relocation information for
  absolute addresses~\cite{retrowrite}.  Nonetheless, this does not
  make symbolization of PIC trivial as we argue in
  Sec.~\ref{sec:symbolization-experiments}.}.That means that for each
original program we test $42$ versions except for
Coreutils where -Ofast generates original binaries that fail many of the tests
and thus we skip it. In summary, we test $3710$ different binaries for
Coreutils, $2898$ binaries for the CGC benchmark, and $1050$ binaries
from our real world selection. All benchmarks together
represent a total of $888$ MB of binaries. Note that
the real world examples represent a significant
portion of the binary data ($324$ MB).

\subsection{Symbolization Experiments}
\label{sec:symbolization-experiments}
We disassemble all the benchmarks
and collect the number of false positives (FP) and false negatives (FN) in the
symbolization procedure. We obtain ground truth by
generating binaries with complete relocation information using the
\texttt{--emit-relocs} ld linker option. We also detect an additional
kind of error WS---i.e. when we create a symbolic expression, but the symbol points to
the wrong section  (see Sec.~\ref{sec:false-negatives}).

\begin{table*}[t]
  \centering
  \begin{tabular}{lr | rrr | rrrr}
   \multirow{2}{*}{Benchmark}   & \multirow{2}{*}{Binaries} & \multicolumn{3}{|c|}{\ddisasm{}} & \multicolumn{4}{|c}{\Ramblr{}}\\
   &  &
        Disasm  & Reassemble & Test  & Disasm & Reassemble & Test & Test w/o ICC \\
    \hline
  Real world   & 1050   & 100.00\% & 100.00\% & 99.90\%  & 99.62\% & 74.29\% & 39.90\% & 45.55\%  \\
  Coreutils    & 3710  & 100.00\% & 100.00\% &100.00\%   &  99.27\% & 88.35\% & 71.26\% & 80.22\% \\
  CGC         & 2865 & 100.00\% & 99.93\% & 99.44\% & 100.00\% & 73.75\% & 51.24\% & 58.18\% \\
  \end{tabular}
\vspace{-7pt}
  \caption{The functionality of binaries reassembled using \ddisasm{}
    and \Ramblr{} as measured using the test suites distributed with the
    binaries.  The  ``Disasm,''
    ``Reassemble,'' and ``Test'' (w/o ICC) columns list the percentage of binaries
    successfully disassembled, reassembled into a new binary, and that
    pass their original test suite (without counting binaries compiled with ICC) respectively.}
  \label{table:functionality}
\end{table*}

For comparison, we run the same experiments using \Ramblr{}, the tool
with the best published symbolization results.
Table~\ref{table:symbolization} contains the results of this
experiment.
Detailed tables with results broken down by compiler and
optimization flag can be found in 
\ifextended
   App.~\ref{sec:detailed_tables}.
\else
   ~\cite{ddisasm-arxiv}.
\fi
The complete set of binaries, detailed
experiment logs, and the scripts to replicate
the experiments can be found at~\cite{ddisasm-artifact}.

\ddisasm{} presents a very low error rate.  This
shows the effectiveness of the approach.  \ddisasm{} builds on many of
the ideas implemented in \Ramblr{}, but makes significant improvements
(see Sec.\ref{sec:discussion}).
App.~\ref{sec:detailed_exps} contains
a discussion of \ddisasm{}'s failures.
\Ramblr{} performs well on Coreutils and CGC compiled with GCC and
Clang (in line with their experiments).
$315$ out of the $323$ broken Coreutils binaries (without counting
ICC) are broken due to a unique symbolization error in the binaries
compiled with Clang 9.0.1.  This illustrates the degree to which
programs in Coreutils share code.
Nonetheless, \Ramblr{}'s precision drops
greatly against the real world examples (39\% of broken examples) and
binaries compiled with ICC (where all optimized binaries are broken).
Additionally, we do not detect WS in \Ramblr{}, as this information is not readily
available. Thus, the numbers in the 'Broken' column are biased against
\ddisasm{} as there might be binaries broken by \Ramblr{} that are not
counted. 

It is worth pointing out that the ground truth extracted
from relocations is incomplete for binaries compiled with ICC.  This
compiler generates jump tables with |Symbol-Symbol| entries. These
jump tables do not need nor have relocations associated to
them---even in PIC.  We believe that this directly contradicts the claim made
by Dinesh et al.~\cite{retrowrite} that x64 PIC code can be symbolized without
heuristics---only using relocations.

The heuristics' weights for both \ibi{} and symbolization have been
manually set and work well generically across compilers and flags.
Importantly, we fixed the weights before running the experiments
on GCC 9.2.1 and Clang 9.0.1. Nonetheless, the results for these two
compilers are on par with the results for the other compilers. Only
5 of a total of 21 broken binaries were compiled with GCC 9.2.1 or Clang
9.0.1.  Thus, the heuristics's
weights are robust across compiler versions.  When ground truth can be
obtained, these weights could be automatically learned and adjusted
based on a program corpus, we leave that for future work.

Finally, we are interested in knowing the importance of different
heuristics. Thus, we repeat the symbolization experiments for the real
world benchmarks deactivating different kinds of heuristics.  We
deactivate heuristics that 1) detect strings, 2) heuristics that use
\dap{}s (``Data access match'' and ``Access conflict''), and 3) both
kinds at the same time. The results are in
Table~\ref{table:symbolization_no_analysis}.  Without both kinds of
heuristics (row 3), we have a high number of FPs. Detecting strings
(row 2) brings this number down, but we miss symbols that look like
strings (FNs). \dap{}s give us additional evidence for those symbols
through the ``Data access match'' heuristic.  With \dap{}s but no
strings (row 1), we also discard some FPs (by detecting objects of
type ``Other'') but not all. The heuristics complement each other.
Note that the 20 FNs produced by \dap{}s correspond to an array of
structs that is correctly detected, but its pointer fields are
accessed with size $4$ instead of $8$ which derails the analysis.

\begin{table}
  \centering
  \begin{tabular}{l|  rrrr}
    Heuristics   &
     \multicolumn{1}{l}{FP} & FN & WS & Broken\\
     \hline
     No Strings   & 59 & 20 & 50 & 53   \\
     No \dap{}  & 45 & 43 & 50 & 49    \\
    No \dap{} \& Strings   & 113 & 0 & 50 & 98\\
  \end{tabular}
  \caption{Symbolization evaluation of \ddisasm{} on the real world benchmarks deactivating groups of heuristics.}
  \label{table:symbolization_no_analysis}
\end{table}

\subsection{Functionality Experiments}
Using the same benchmarks we check how many of the disassembled
binaries can be reassembled and how many of those pass their original
test suites without errors.

For \ddisasm{}, we perform the experiment on the stripped versions of
the binaries. Additionally, in order to increase our confidence that
both \ibi{} and symbolization are correct, we modify the locations
(and relative locations) of all the instructions by adding NOPs at
regular intervals before reassembling. We add 8 NOPs every 8
instructions to maintain the original instructions' alignment
throughout the executable section\footnote{We skip regions in between
  jump table entries of the form \lstinline{.byte
    Symbol-Symbol}. Adding NOPs to these regions can easily make the
  result of \lstinline{Symbol-Symbol} fall out of the
  range expressible with one byte.}. We also add 64 zero bytes at the beginning of
each data section.  This demonstrates that our symbolization is robust
to significant modification of
code (by adding or removing code) and data (by adding content
at the beginning of sections).

For \Ramblr{}, we use unstripped binaries because \Ramblr{} fails to
produce reassembleable assembly for the stripped versions of most
binaries. Many of the failures are because \Ramblr{} generates
assembly with undefined labels or with labels defined
twice. This kind of inconsistency is easy to avoid
in a Datalog implementation.
Additionally, we do not perform any modification of assembly
generated by \Ramblr{}---this ensures that we do not report an overly
pessimistic result for \Ramblr{} by accidentally breaking the code
generated by \Ramblr{}. So we compare \ddisasm{} at a significant
handicap against \Ramblr{}.

The results of this experiment are in Table~\ref{table:functionality}.
For CGC, we discarded 33 binaries that fail their tests
non-deterministically leaving 2865 binaries.  \ddisasm{} produces
reassembleable assembly code for all the binaries but two.  One binary
in the real world benchmarks and $14$ binaries in the CGC benchmark
fail their tests. This is close to the results of our previous
experiment (Table~\ref{table:symbolization}).  The FNs in real world
examples and 5 of the 17 FNs in CGC cause 1 and 5 test failures respectively.
The remaining
FPs, FNs, and WS symbols do not cause test failures.  Additionally,
there are 9 other test failures in CGC not caused by symbolization
errors.

\begin{figure}[t]
  \ifarxiv
    \includegraphics{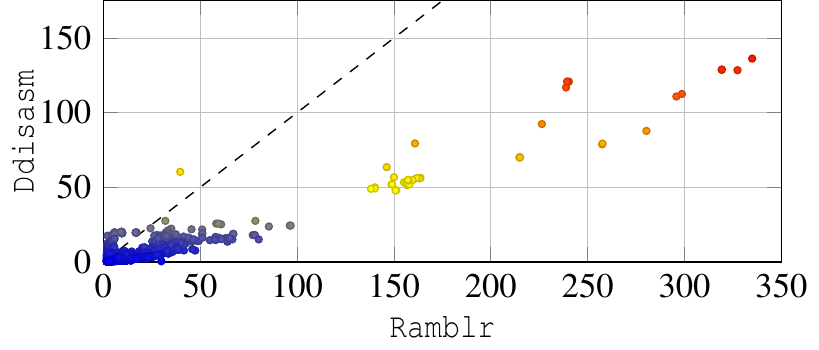}
  \else
    \begin{tikzpicture} \begin{axis}[
        xmin=0,
        ymin=0,
        height=0.5\linewidth,
        width=\linewidth,
        xmax=350,
        ymax=175,
        mark size=1pt,
        xlabel=\Ramblr{},
        ylabel=\ddisasm{},
        y label style={at={(0.06,0.5)}},
        x label style={at={(0.5,0.04)}},
        grid=major,
        major grid style={line width=.2pt,draw=gray!50}
      ]
      \addplot[scatter,only marks]
       table[ x index=16,y index=5, col sep=tab]{perf_data/joined_cgc.data};
       \addplot[scatter,only marks]
      table[ x index=16,y index=5, col sep=tab]{perf_data/joined_real.data};
      \addplot[scatter,only marks]
      table[ x index=15,y index=5, col sep=tab]{perf_data/joined_coreutils.data};
      
      \draw [dashed] (rel axis cs:0,0) -- (rel axis cs:0.5,1);
    \end{axis}
    \end{tikzpicture}
  \fi
  \ifarxiv
    \includegraphics{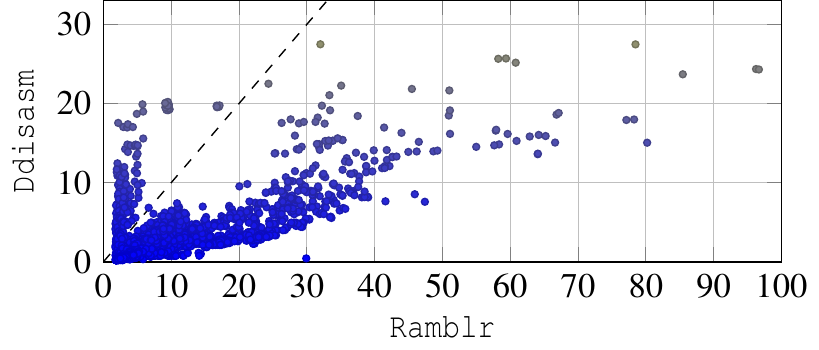}
  \else
  \begin{tikzpicture} \begin{axis}[
        xmin=0,
        ymin=0,
        height=0.5\linewidth,
        width=\linewidth,
        xmax=100,
        ymax=33,
        xtick distance=10,
        ytick distance=10,
        mark size=1pt,
        xlabel=\Ramblr{},
        ylabel=\ddisasm{},
        y label style={at={(0.06,0.5)}},
        x label style={at={(0.5,0.04)}},
        grid=major,
        major grid style={line width=.2pt,draw=gray!50}
      ]
      \addplot[scatter,only marks]
      table[ x index=16,y index=5, col sep=tab]{perf_data/joined_real.data};
      \addplot[scatter,only marks]
      table[ x index=15,y index=5, col sep=tab]{perf_data/joined_coreutils.data};
       \addplot[scatter,only marks]
      table[ x index=16,y index=5, col sep=tab]{perf_data/joined_cgc.data};
      \draw [dashed] (rel axis cs:0,0) -- (rel axis cs:0.33,1);
    \end{axis}
  \end{tikzpicture}
  \fi
     \caption{Disassembly time.  The two
         graphs show the disassembly times (in seconds) for all the
         binaries at two different scales (the bottom graph displays
         smaller binaries in detail).   \ddisasm{}'s disassembly
         time is plotted
         (vertically) against  \Ramblr{}'s
         (horizontally).  In all graphs, points below the diagonal
         represent binaries for which \ddisasm{} is faster than
         \Ramblr{}.}
    \label{fig:performance}
\end{figure}

\subsection{Performance Evaluation}
Finally, we measure and compare the performance of both \Ramblr{} and
\ddisasm{}. We measure the time that it takes to disassemble each of the
binaries in the three benchmarks. The results can be found in
Fig.~\ref{fig:performance}.  \ddisasm{} is faster than \Ramblr{} in all
but 294 of 7658 total binaries. In particular, \ddisasm{} is on average
 4.9 times faster than \Ramblr{}.

\section{Conclusion}

We have developed a new reassembleable disassembler called \ddisasm{}.
\ddisasm{} in implemented in Datalog and combines novel static analyses
and heuristics to determine how data is accessed and used.  We show
that Datalog is well suited to this task as it enables the
compositional and declarative specification of static analyses and
heuristics, and it compiles them into a unified, parallel, and
efficient executable.

\ddisasm{} is, to the best of our knowledge, the first disassembler for
machine code implemented in Datalog. Our experiments show that
\ddisasm{} is both more precise and faster than the state-of-the-art
tools for reassembleable disassembly, and better handles large complex
real-world programs.  \ddisasm{} makes binary rewriting practical by
enabling binary rewriting of real world programs compiled with a range
of compilers and optimization levels with unprecedented speed and
accuracy.

\ifanonymous
\else
\section{Acknowledgments}

This material is based upon work supported by the Office of Naval
Research under contract No. N68335-17-C-0700. Any opinions, findings
and conclusions or recommendations expressed in this material are
those of the authors and do not necessarily reflect the views of the
Office of Naval Research.

\fi

\bibliographystyle{plain}
\bibliography{biblio}

\appendix

\section{Symbol-Symbol Jump Tables}
\label{sec:jump-tables}
This appendix describes jump tables with relative offsets and how they
are detected by our disassembler.

Most jump tables in programs compiled with GCC and Clang (position
dependent code) are lists of absolute addresses that can be detected
like any other symbolic value. This is not the case for jump tables
generated by ICC and jump tables generated by PIC code.  These jump
tables are often expressed as lists of |Symbol-Symbol| expressions.

In this kind of jump tables, one of the symbols represents a reference
point, and the other symbol represents the jump target. The reference
point is the same for all the jump table entries and the actual value
stored at each the jump table entry is the distance between the jump
target and the reference point. The size of jump table entries can
vary i.e.  $1$, $2$, $4$ or $8$ bytes.

\begin{figure}[h]
  \lstset{language=[x64]Assembler}
  \begin{lstlisting}
47DA7b:   lea RDX, QWORD PTR [RIP+.L_4A09F0]
47DA82:   movzx EDX, BYTE PTR [RDX+RCX*1]
47DA86:   lea RAX, QWORD PTR [RIP+.L_47DA93]
47DA8d:   add RAX,RDX
47DA90:   jmp RAX
 
4a09f0:   .byte .L_47DB3F-.L_47DA93
          .byte .L_47DB36-.L_47DA93
          .byte .L_47DB2B-.L_47DA93
          .byte .L_47DB20-.L_47DA93
\end{lstlisting}
\caption{Assembly  (after symbolization) extracted from tar-1.29 compiled with ICC
  -O2. This code implements a jump table of \lstinline{Symbol-Symbol} entries
  of size $1$ byte.}
\label{fig:jump-table}
\end{figure}

\begin{example}
  Consider the example code in Fig.~\ref{fig:jump-table}.  The first
  instruction loads the start address of the jump table onto |RDX|;
  the second instruction reads the jump table entry and stores it in |RDX|; the third
  instruction loads address |47DA93| that acts as a reference for the jump
  table onto |RAX|; the fourth instruction computes the jump
  target by adding |RAX| and |RDX| and the last instruction executes
  the jump.
\end{example}

In order to find a jump table, we need to determine: the jump table starting point,
the jump table reference point, and the size of each jump table entry.
Fortunately, the code patterns used to implement this kind of jump
tables are relatively regular. We have specialized Datalog rules to detect them.

\begin{drule}{drule:jumptable}
\begin{lstlisting}
jump_table_start(AJump,Size,Start,Reference):-
    reg_jump(AJump,_),
    def_used(ASum,Reg,AJump,_),
    reg_reg_op(ASum,Reg,RegEntry,RegRef,1,0),
    
    def_used(AEntry,RegEntry,ASum,_),
    data_access_pattern(Start,Size,Size,AEntry),

    def_used(ARef,RegRef,ASum,_),
    reg_val(ARef,RegRef,_,'NONE',0,Reference).
\end{lstlisting}
\end{drule}

Rule~\ref{drule:jumptable} is simplified version of the rule that
detect the pattern in Fig.~\ref{fig:jump-table}. The rule finds a jump
that uses a register and ``walks back'' the code using def-use chains to the
instruction where the jump target is computed (at address |Asum|).  At
that location, |reg_reg_op| represents an abstraction of an assembly
instruction on two registers |Reg=RegEntry+RegRef$\times$1+0|.  Then, the
rule examines the definition of |RegEntry| to find where the jump
table entry is read (at address |AEntry|) and thanks to its
|data_access_pattern|, it determines the jump table starting
address |Start| and the size of each entry |Size|.  The other register
|RegRef| should contain the jump table reference point. So its value is
obtained using |reg_val| which should contain a constant value (not
expressed in terms of another register).

By relying on the analyses presented in Sec.~\ref{sec:analysis}, i.e
def-use chains, \dap{}s and the register value analysis;
the Datalog rule is more robust than exact pattern matching. The
instructions involved in the jump table do not necessarily appear all
together or in a fixed order, and the rule does not rely on specific
instructions being used. E.g. the jump target computation is sometimes done using
|LEA| instead of |ADD|.

Once we have found the jump table beginning and its corresponding
|data_access_pattern|, we can use the |propagated_data_access| (see
Sec.~\ref{sec:data-access}) to create |symbol-symbol| candidates for
each of the jump table entries. That means that we will consider that
the jump table extends until there is another data access from a
different part of the code.

The detection of these jump tables has been the main addition required
to support the ICC compiler. Other analyses and heuristics have
remained largely the same. We expect that supporting additional
compilers will require similar additions as each compiler has its own
particular code patterns. However, the analyses described in
Sec.~\ref{sec:analysis} remain useful building blocks that facilitate
supporting these special constructs in a robust manner.

\section{Symbolization Failures}
\label{sec:detailed_exps}

We manually examined and diagnosed \ddisasm{}'s
symbolization failures to determine what are the causes that lead
to the remaining FPs, FNs or WS.

\paragraph*{Real world Benchmarks} In
the real world benchmarks, the $20$ FPs corresponds to a
single array of structs that contains pointers. Our analysis obtains
the right \dap{} with the right multipliers but only $4$ bytes of each
of the pointers are read instead of $8$. This leads \ddisasm{} to
conclude that those locations contains data objects of type ``Other''
of size $4$ instead of symbols. These FNs cause the corresponding tests
to fail.

The $50$ symbols pointing to the wrong section (WS) are displacements in
indirect operands and happen in $5$ variants of the same program (lighttpd-1.4.18)
compiled with Clang 6.0 and Clang 9.0.1. These particular cases are
not currently detected by our heuristics but they also do not cause
test failures in our functionality experiments.

\paragraph*{Coreutils Benchmarks}
In Coreutils, there are $3$ FPs, all in binaries compiled
with -O0. They correspond to immediate operands that are moved or
compared to registers. Those registers are loaded from the stack
immediately before the location of the immediate and they are stored
in the stack again immediately after. Therefore, our analyses do not
obtain any evidence on the type of those immediates.  These FPs do not
cause tests failures, probably because the Coreutils test suites are not
exhaustive.

\paragraph*{CGC Benchmarks}
In the CGC benchmarks, $5$ of the $12$ ``Broken'' binaries have FNs
where the corresponding relocations refer to the symbols
|__init_array_start| and |__init_array_end|. These binaries, compiled
with ICC, do not have an |.init_array| section and in fact the
symbols' addresses are the same and fall outside all data
sections. Nonetheless, the code uses the difference between the two symbols (which
is zero) and thus it has the same behavior even though these
references have not been made symbolic. In fact, we do not observe
test failures in these binaries.

There are $2$ other binaries, variants of the same program compiled
with ICC, that have displacements in an indirect operand pointing to the
wrong section (WS). These particular cases are not currently detected by our
heuristics. They also do not cause test failures.

Two variants of the same binary compiled with Clang 9.0.1 have a FN in
an indirect operand. The symbol candidate points to the end of the
|.rodata| section which coincides with the beginning of
|.eh_frame_hdr|. This triggers the ``Pointer to special section''
heuristic which leads \ddisasm{} to incorrectly discard the symbol candidate.  We
plan to refine the ``Pointer to special section'' heuristic to avoid
this corner case.

The $3$ remaining failures are due to FN in variants of
the same program compiled with GCC 7.1. They correspond to an
immediate that should be a |Symbol+Constant|. The immediate is a loop
bound but it corresponds to a triple nested loop that our heuristics
do not detect well. The extended section considered is not large
enough for the constant required by the immediate.
These FPs cause the tests to fail.

\ifextended
 \section{Detailed Experimental Results}
\label{sec:detailed_tables}

Tables~\ref{table:symbolization_detail_rw},
\ref{table:symbolization_detail_core}, and
\ref{table:symbolization_detail_cgc} contain the results of the
symbolization broken down by compiler and optimization flag.
Tables~\ref{table:functionality_detail_rw},
\ref{table:functionality_detail_core}, and
\ref{table:functionality_detail_cgc} contain the results of the
functionality experiments broken down by compiler and optimization
flag. These tables present absolute numbers rather than percentages to
facilitate a more detailed analysis.

\begin{table*}[hbt]
  \centering
  \begin{tabular}{llll|  rrrr|  rrr}
    \hline
      \multicolumn{10}{c}{Real World Benchmarks} \\ \hline
     \multirow{2}{*}{Compiler}& \multirow{2}{*}{Optimization} & \multirow{2}{*}{Binaries} &
    \multirow{2}{*}{References} &
     \multicolumn{4}{c}{\ddisasm{}} & \multicolumn{3}{|c}{\Ramblr{}}\\
    &&&& \multicolumn{1}{l}{FP} & FN & WS & Broken  & FP & FN  & Broken\\
     \hline
\multirow{6}{*}{GCC 5.5} & -O0 & 25 &146169 & 0 & 0 & 0 & 0 &            50 & 8  & 7 \\       
                     & -O1 & 25 &139982 & 0 & 0 & 0 & 0 &            50 & 8  & 8 \\           
                     & -O2 & 25 &136633 & 0 & 0 & 0 & 0 &            49 & 8  & 8 \\    
                     & -O3 & 25 &162674 & 0 & 0 & 0 & 0 &            53 & 9  & 9 \\            
                  & -Ofast & 25 &162664 & 0 & 0 & 0 & 0 &            49 & 9  & 8 \\              
                     & -Os & 25 &118192 & 0 & 0 & 0 & 0 &            42 & 9  & 7 \\             
\hline                                                                                      
\multirow{6}{*}{GCC 7.1} & -O0 & 25& 146825 & 0 & 0 & 0 & 0 &          50 & 8  & 7 \\       
                      & -O1 & 25 & 139653 & 0 & 0 & 0 & 0 &          50 & 9  & 8 \\   
                      & -O2 & 25 & 136776 & 0 & 0 & 0 & 0 &          49 & 8  & 8 \\   
                      & -O3 & 25 & 165484 & 0 & 0 & 0 & 0 &          55 & 48 & 8 \\  
                   & -Ofast & 25 & 165477 & 0 & 0 & 0 & 0 &          52 & 43 & 8 \\  
                      & -Os & 25 & 117889 & 0 & 0 & 0 & 0 &          42 & 9  & 7 \\   
\hline                                                                                       
\multirow{6}{*}{Clang 3.8} & -O0 & 25& 132401 & 0 & 0 & 0 & 0 &          39 & 9  & 6 \\       
                      & -O1 & 25 & 125023 & 0 & 0 & 0 & 0 &          39 & 12  & 8 \\  
                      & -O2 & 25 & 142910 & 0 & 0 & 0 & 0 &          41 & 20  & 9 \\  
                      & -O3 & 25 & 149566 & 0 & 0 & 0 & 0 &          43 & 21  & 9 \\  
                   & -Ofast & 25 & 149539 & 0 & 0 & 0 & 0 &          34 & 20  & 8 \\
                      & -Os & 25 & 126359 & 0 & 0 & 0 & 0 &          38 & 11  & 8 \\
\hline                                                                                       
\multirow{6}{*}{Clang 6.0}& -O0 & 25 & 142370 & 0 & 0 & 0 & 0 &      38 & 8  & 6 \\       
                          & -O1 & 25 & 134185 & 0 & 0 & 0 & 0 &      42 & 12  & 8 \\
                          & -O2 & 25 & 157599 & 0 & 0 & 0 & 0 &      41 & 25  & 8 \\
                          & -O3 & 25 & 164270 & 0 & 0 & 10 & 1 &     41 & 20  & 8 \\
                       & -Ofast & 25 & 164277 & 0 & 0 & 10 & 1 &     38 & 22  & 8 \\
                          & -Os & 25 & 124989 & 0 & 0 & 0 & 0 &      41 & 12  & 8 \\
\hline                                                                                       
\multirow{6}{*}{ICC}  & -O0 & 25 & 152957 & 0 & 0 & 0 & 0 &          3044 & 2403  & 10 \\   
                      & -O1 & 25 & 121901 & 0 & 20 & 0 & 1 &         8762 & 3832  & 25 \\
                      & -O2 & 25 & 219130 & 0 & 0 & 0 & 0 &          9133 & 16529  & 25 \\
                      & -O3 & 25 & 227835 & 0 & 0 & 0 & 0 &          9491 & 17460  & 25 \\
                      & -Ofast & 25 & 227832 & 0 & 0 & 0 & 0 &          9489 & 17462  & 25 \\
                      & -Os & 25 & 119754 & 0 & 0 & 0 & 0 &          8754 & 3829  & 25 \\
\hline
\multirow{6}{*}{GCC 9.2.1} & -O0 & 25   &144719  & 0 & 0 & 0 & 0 &      44 & 8 & 6 \\ 
 & -O1 & 25                             &138365  & 0 & 0 & 0 & 0 &     49 & 8 & 7 \\ 
 & -O2 & 25                             &125880  & 0 & 0 & 0 & 0 &     48 & 39 & 8 \\
 & -O3 & 25                             &139479  & 0 & 0 & 0 & 0 &  50 & 9 & 9 \\ 
 & -Ofast & 25                          &139491  & 0 & 0 & 0 & 0 &   46 & 9 & 8 \\ 
 & -Os & 25                             &120956  & 0 & 0 & 0 & 0 &     42 & 10 & 7 \\
\hline                                                                                
\multirow{6}{*}{Clang 9.0.1} & -O0 & 25 &99843   & 0 & 0 & 0 & 0 &   38 & 8 & 6 \\  
 & -O1 & 25                             &96776   & 0 & 0 & 0 & 0 &  40 & 11 & 7 \\ 
 & -O2 & 25                             &111642  & 0 & 0 & 10 & 1 &   41 & 20 & 7 \\
 & -O3 & 25                             &115425  & 0 & 0 & 10 & 1 &  42 & 21 & 7 \\
 & -Ofast & 25                          &115432  & 0 & 0 & 10 & 1 &  38 & 22 & 7 \\
 & -Os & 25                             &87693   & 0 & 0 & 0 & 0 &   41 & 12 & 7 \\ 
\hline
  \end{tabular}
 
  \caption{Real world benchmark symbolization evaluation. Results broken down by compiler and optimization flag.}
  \label{table:symbolization_detail_rw}
\end{table*}

\begin{table*}[hbt]
  \centering
  \begin{tabular}{llll|  rrrr|  rrr}
    \hline
      \multicolumn{10}{c}{Coreutils Benchmarks} \\ \hline
     \multirow{2}{*}{Compiler}& \multirow{2}{*}{Optimization} & \multirow{2}{*}{Binaries} &
    \multirow{2}{*}{References} &
     \multicolumn{4}{c}{\ddisasm{}} & \multicolumn{3}{|c}{\Ramblr{}}\\
    &&&& \multicolumn{1}{l}{FP} & FN & WS & Broken  & FP & FN  & Broken\\
     \hline
\multirow{5}{*}{GCC 5.5} & -O0 & 106   & 124991 & 0 & 0 & 0 & 0 &  0 & 0  & 0 \\        
                       & -O1 & 106 & 123239 & 0 & 0 & 0 & 0 &  0 & 0  & 0 \\        
                       & -O2 & 106 & 113264 & 0 & 0 & 0 & 0 &  0 & 0  & 0 \\        
                       & -O3 & 106 & 171177 & 0 & 0 & 0 & 0 &  3 & 0  & 2 \\        
                       & -Os & 106 & 81661  & 0 & 0 & 0 & 0 & 0 & 0   & 0 \\         
\hline                                                                                 
\multirow{5}{*}{GCC 7.1} & -O0 & 106 & 124919 & 0 & 0 & 0 & 0 &  0 & 0  & 0 \\        
                       & -O1 & 106 & 123205 & 0 & 0 & 0 & 0 &  0 & 0  & 0 \\        
                       & -O2 & 106 & 122142 & 0 & 0 & 0 & 0 &  1 & 0  & 1 \\        
                       & -O3 & 106 & 188624 & 0 & 0 & 0 & 0 &  2 & 0  & 2 \\        
                       & -Os & 106 & 81319  & 0 & 0 & 0 & 0 & 0 & 0  & 0 \\         
\hline                                                                                 
\multirow{5}{*}{Clang 3.8} & -O0 & 106 & 98876  & 0 & 0 & 0 & 0 & 0 & 0 & 0 \\         
                       & -O1 & 106 & 97662  & 0 & 0 & 0 & 0 & 0 & 0 & 0 \\         
                       & -O2 & 106 & 107051 & 0 & 0 & 0 & 0 &  0 & 0  & 0 \\        
                       & -O3 & 106 & 108836 & 0 & 0 & 0 & 0 &  0 & 0  & 0 \\        
                       & -Os & 106 & 98669  & 0 & 0 & 0 & 0 & 0 & 0  & 0 \\         
\hline                                                                                 
\multirow{5}{*}{Clang 6.0}&-O0&106 & 116987 & 1 & 0 & 0 & 1 &  1 & 0  & 1 \\        
                       & -O1 & 106 & 118466 & 0 & 0 & 0 & 0 &  0 & 0  & 0 \\        
                       & -O2 & 106 & 118200 & 0 & 0 & 0 & 0 &  0 & 0  & 0 \\        
                       & -O3 & 106 & 123213 & 0 & 0 & 0 & 0 &  0 & 0  & 0 \\        
                       & -Os & 106 & 98565  & 0 & 0 & 0 & 0 & 0 & 0  & 0 \\         
\hline                                                                                 
\multirow{5}{*}{ICC} & -O0 & 106   & 112111 & 2 & 0 & 0 & 2 & 3 & 0  & 3 \\         
                       & -O1 & 106 & 104858 & 0 & 0 & 0 & 0 & 455 & 6572  & 106 \\  
                       & -O2 & 106 & 270000 & 0 & 0 & 0 & 0 & 3662 & 61709  & 106 \\
                       & -O3 & 106 & 272180 & 0 & 0 & 0 & 0 & 3662 & 62141  & 106 \\
                       & -Os & 106 & 104657 & 0 & 0 & 0 & 0 & 455 & 6572  & 106 \\  
\hline
\multirow{5}{*}{GCC 9.2.1} & -O0 & 106   & 116252 & 0 & 0 & 0 & 0 & 0 & 0 & 0 \\    
 & -O1 & 106                         & 146982 & 0 & 0 & 0 & 0 &  0 & 0 & 0 \\    
 & -O2 & 106                         & 144085 & 0 & 0 & 0 & 0 &  1 & 0 & 1 \\    
 & -O3 & 106                         & 196210 & 0 & 0 & 0 & 0 &  1 & 0 & 1 \\    
 & -Os & 106                         & 95336  & 0 & 0 & 0 & 0 &  0 & 0 & 0 \\     
\hline                                                                              
\multirow{5}{*}{Clang 9.0.1} & -O0 & 106 & 74666  & 0 & 0 & 0 & 0 &  0 & 0 & 0 \\     
 & -O1 & 106                         & 80690  & 0 & 0 & 0 & 0 & 0 & 1260 & 105 \\
 & -O2 & 106                         & 78068  & 0 & 0 & 0 & 0 & 0 & 1260 & 105 \\
 & -O3 & 106                         & 81567  & 0 & 0 & 0 & 0 & 0 & 1260 & 105 \\
 & -Os & 106                         & 60611  & 0 & 0 & 0 & 0 & 0 & 0 & 0 \\     
\hline
\end{tabular}
 
  \caption{Coreutils benchmark symbolization evaluation. Results broken down by compiler and optimization flag.}
  \label{table:symbolization_detail_core}
\end{table*}

\begin{table*}[hbt]
  \centering
  \begin{tabular}{llll|  rrrr|  rrr}
    \hline
      \multicolumn{10}{c}{CGC Benchmarks} \\ \hline
     \multirow{2}{*}{Compiler}& \multirow{2}{*}{Optimization} & \multirow{2}{*}{Binaries} &
    \multirow{2}{*}{References} &
     \multicolumn{4}{c}{\ddisasm{}} & \multicolumn{3}{|c}{\Ramblr{}}\\
    &&&& \multicolumn{1}{l}{FP} & FN & WS & Broken  & FP & FN  & Broken\\
     \hline
\multirow{6}{*}{GCC 5.5} & -O0 & 69   & 281509 & 0 & 0 & 0 & 0 & 0 & 0 & 0 \\       
                       & -O1 & 69 & 278277 & 0 & 0 & 0 & 0 & 0 & 1 & 1 \\       
                       & -O2 & 69 & 130326 & 0 & 0 & 0 & 0 & 1 & 1 & 2 \\       
                       & -O3 & 69 & 135812 & 0 & 0 & 0 & 0 & 3 & 1 & 4 \\       
                    & -Ofast & 69 & 135759 & 0 & 0 & 0 & 0 & 3 & 1 & 4 \\       
                       & -Os & 69 & 127048 & 0 & 0 & 0 & 0 & 0 & 0 & 0 \\       
\hline                                                                          
\multirow{6}{*}{GCC 7.1} & -O0 & 69 & 281452 & 0 & 0 & 0 & 0 & 0 & 0 & 0 \\       
                       & -O1 & 69 & 278275 & 0 & 0 & 0 & 0 & 0 & 1 & 1 \\       
                       & -O2 & 69 & 130253 & 0 & 1 & 0 & 1 & 1 & 1 & 2 \\       
                       & -O3 & 69 & 136674 & 0 & 1 & 0 & 1 & 2 & 1 & 3 \\       
                    & -Ofast & 69 & 136658 & 0 & 1 & 0 & 1 & 2 & 1 & 3 \\       
                       & -Os & 69 & 127088 & 0 & 0 & 0 & 0 & 0 & 0 & 0 \\       
\hline                                                                          
\multirow{6}{*}{Clang 3.8} & -O0 & 69 & 297970 & 0 & 0 & 0 & 0 & 0 & 0 & 0 \\       
                       & -O1 & 69 & 231882 & 0 & 0 & 1 & 1 & 1 & 0 & 1 \\       
                       & -O2 & 69 & 235458 & 0 & 0 & 0 & 0 & 1 & 0 & 1 \\       
                       & -O3 & 69 & 236308 & 0 & 0 & 0 & 0 & 1 & 0 & 1 \\       
                    & -Ofast & 69 & 236319 & 0 & 0 & 0 & 0 & 1 & 0 & 1 \\       
                       & -Os & 69 & 128936 & 0 & 0 & 0 & 0 & 0 & 0 & 0 \\       
\hline                                                                          
\multirow{6}{*}{Clang 6.0}&-O0&69 & 298708 & 0 & 0 & 0 & 0 & 0 & 0 & 0 \\       
                       & -O1 & 69 & 135007 & 0 & 0 & 1 & 1 & 1 & 0 & 1 \\       
                       & -O2 & 69 & 140984 & 0 & 0 & 0 & 0 & 1 & 0 & 1 \\       
                       & -O3 & 69 & 141717 & 0 & 0 & 0 & 0 & 1 & 0 & 1 \\       
                    & -Ofast & 69 & 141744 & 0 & 0 & 0 & 0 & 1 & 0 & 1 \\       
                       & -Os & 69 & 128649 & 0 & 0 & 0 & 0 & 0 & 0 & 0 \\       
\hline                                                                          
\multirow{6}{*}{ICC} & -O0 & 69   & 287675 & 0 & 0 & 0 & 0 & 150 & 2160 & 15 \\ 
                       & -O1 & 69 & 150848 & 0 & 4 & 0 & 2 & 2123 & 6406 & 69 \\
                       & -O2 & 69 & 234370 & 0 & 0 & 0 & 0 & 2157 & 9561 & 69 \\
                       & -O3 & 69 & 236434 & 0 & 3 & 0 & 1 & 2158 & 9572 & 69 \\
                    & -Ofast & 69 & 236442 & 0 & 3 & 0 & 1 & 2158 & 9572 & 69 \\
                       & -Os & 69 & 151004 & 0 & 2 & 0 & 1 & 2123 & 6404 & 69 \\
\hline

\multirow{6}{*}{GCC 9.2.1} & -O0 & 69   &282186 & 0 & 0 & 0 & 0 & 0 & 0 & 0 \\
 & -O1 & 69                         &277899 & 0 & 0 & 0 & 0 & 0 & 0 & 0 \\
 & -O2 & 69                         &130989 & 0 & 0 & 0 & 0 & 1 & 0 & 1 \\
 & -O3 & 69                         &137572 & 0 & 0 & 0 & 0 & 1 & 0 & 1 \\
 & -Ofast & 69                      &137548 & 0 & 0 & 0 & 0 & 1 & 0 & 1 \\
 & -Os & 69                         &127108 & 0 & 0 & 0 & 0 & 0 & 0 & 0 \\
\hline                                                                    
\multirow{6}{*}{Clang 9.0.1} & -O0 & 69 &70557  & 0 & 0 & 0 & 0 & 0 & 0 & 0 \\
 & -O1 & 69                         &44242  & 0 & 1 & 0 & 1 & 0 & 0 & 0 \\
 & -O2 & 69                         &48272  & 0 & 0 & 0 & 0 & 0 & 0 & 0 \\
 & -O3 & 69                         &48343  & 0 & 0 & 0 & 0 & 0 & 0 & 0 \\
 & -Ofast & 69                      &48367  & 0 & 0 & 0 & 0 & 0 & 0 & 0 \\
 & -Os & 69                         &37782  & 0 & 1 & 0 & 1 & 0 & 0 & 0 \\
\hline
  \end{tabular}
 
  \caption{CGC benchmark symbolization evaluation. Results broken down by compiler and optimization flag.}
  \label{table:symbolization_detail_cgc}
\end{table*}

\begin{table*}[t]
  \centering
  \begin{tabular}{llr | rrr | rrr}
  \hline
  \multicolumn{9}{c}{Real World Benchmarks} \\ \hline
   \multirow{2}{*}{Compiler}& \multirow{2}{*}{Optimization} & \multirow{2}{*}{Binaries} & \multicolumn{3}{|c|}{\ddisasm{}} & \multicolumn{3}{|c}{\Ramblr{}}\\
   &  & &  Disasm  & Reassemble & Test  & Disasm & Reassemble & Test \\
    \hline

    \multirow{6}{*}{GCC 5.5} & -O0& 25& 25& 25& 25& 25 & 21 & 19 \\
 & -O1 & 25 & 25 & 25 & 25 &                        25 & 21 & 16 \\
 & -O2 & 25 & 25 & 25 & 25 &                        25 & 21 & 16 \\
 & -O3 & 25 & 25 & 25 & 25 &                        25 & 20 & 9 \\  
 & -Ofast & 25 & 25 & 25 & 25 &                     25 & 21 & 11 \\ 
 & -Os & 25 & 25 & 25 & 25 &                        25 & 21 & 14 \\
\hline                                                             
\multirow{6}{*}{GCC 7.1} & -O0 & 25 & 25 & 25& 25&  25 & 21 & 14 \\
 & -O1 & 25 & 25 & 25 & 25 &                        25 & 21 & 15 \\
 & -O2 & 25 & 25 & 25 & 25 &                        25 & 21 & 11 \\
 & -O3 & 25 & 25 & 25 & 25 &                        25 & 21 & 6 \\  
 & -Ofast & 25 & 25 & 25 & 25 &                     25 & 21 & 6 \\  
 & -Os & 25 & 25 & 25 & 25 &                        25 & 21 & 11 \\
\hline                                                             
\multirow{6}{*}{Clang 3.8}&-O0 & 25 & 25 & 25 & 25& 25 & 22 & 17 \\
 & -O1 & 25 & 25 & 25 & 25 &                        25 & 21 & 11 \\
 & -O2 & 25 & 25 & 25 & 25 &                        25 & 21 & 10 \\
 & -O3 & 25 & 25 & 25 & 25 &                        25 & 21 & 10 \\
 & -Ofast & 25 & 25 & 25 & 25 &                     25 & 23 & 12 \\
 & -Os & 25 & 25 & 25 & 25 &                        25 & 21 & 9 \\  
\hline                                                             
\multirow{6}{*}{Clang 6.0}& -O0 & 25& 25& 25& 25&   25 & 22 & 17 \\                     
 & -O1 & 25 & 25 & 25 & 25 &                        25 & 21 & 10 \\
 & -O2 & 25 & 25 & 25 & 25 &                        25 & 21 & 8 \\  
 & -O3 & 25 & 25 & 25 & 25 &                        25 & 21 & 9 \\  
 & -Ofast & 25 & 25 & 25 & 25 &                     25 & 23 & 11 \\ 
 & -Os & 25 & 25 & 25 & 25 &                        25 & 21 & 9 \\  
\hline                                                             
\multirow{6}{*}{ICC} & -O0 & 25 & 25 & 25 & 25 &   25 & 17 & 9 \\ 
 & -O1 & 25 & 25 & 25 & 24 &                       25 & 0 & 0 \\  
 & -O2 & 25 & 25 & 25 & 25 &                       24 & 0 & 0 \\  
 & -O3 & 25 & 25 & 25 & 25 &                       25 & 0 & 0 \\  
 & -Ofast & 25 & 25 & 25 & 25 &                    25 & 0 & 0 \\  
 & -Os & 25 & 25 & 25 & 25 &                       25 & 0 & 0 \\  
\hline
\multirow{6}{*}{GCC 9.2.1} & -O0 & 25 & 25 & 25 & 25 &   25 & 21 & 14 \\
 & -O1 & 25 & 25 & 25 & 25 &                         25 & 21 & 14 \\
 & -O2 & 25 & 25 & 25 & 25 &                         24 & 20 & 11 \\
 & -O3 & 25 & 25 & 25 & 25 &                         24 & 20 & 7 \\ 
 & -Ofast & 25 & 25 & 25 & 25 &                      24 & 22 & 8 \\ 
 & -Os & 25 & 25 & 25 & 25 &                         25 & 21 & 13 \\
\hline                                                              
\multirow{6}{*}{Clang 9.0.1} & -O0 & 25 & 25 & 25 & 25 & 25 & 22 & 16 \\
 & -O1 & 25 & 25 & 25 & 25 &                         25 & 21 & 10 \\
 & -O2 & 25 & 25 & 25 & 25 &                         25 & 21 & 9 \\ 
 & -O3 & 25 & 25 & 25 & 25 &                         25 & 21 & 9 \\ 
 & -Ofast & 25 & 25 & 25 & 25 &                      25 & 23 & 9 \\ 
 & -Os & 25 & 25 & 25 & 25 &                         25 & 21 & 9 \\ 
\hline
  \end{tabular}
  \caption{Real World benchmarks functionality evaluation. Results broken down by compiler and optimization flag.}
  \label{table:functionality_detail_rw}
\end{table*}

\begin{table*}[t]
  \centering
  \begin{tabular}{llr | rrr | rrr}
  \hline
  \multicolumn{9}{c}{Coretuils Benchmarks} \\ \hline
   \multirow{2}{*}{Compiler}& \multirow{2}{*}{Optimization} & \multirow{2}{*}{Binaries} & \multicolumn{3}{|c|}{\ddisasm{}} & \multicolumn{3}{|c}{\Ramblr{}}\\
   &  & &  Disasm  & Reassemble & Test  & Disasm & Reassemble & Test \\
    \hline
\multirow{5}{*}{GCC 5.5} & -O0 & 106 & 106 & 106 & 106      &  106 & 106 & 93 \\
 & -O1 & 106 & 106 & 106 & 106 &                               106 & 106 & 93 \\
 & -O2 & 106 & 106 & 106 & 106 &                               106 & 105 & 92 \\
 & -O3 & 106 & 106 & 106 & 106 &                               106 & 104 & 87 \\
 & -Os & 106 & 106 & 106 & 106 &                               106 & 106 & 78 \\
\hline                                                                          
\multirow{5}{*}{GCC 7.1} & -O0 & 106 & 106 & 106 & 106 &       106 & 106 & 93 \\
 & -O1 & 106 & 106 & 106 & 106 &                               106 & 106 & 92 \\
 & -O2 & 106 & 106 & 106 & 106 &                               106 & 105 & 76 \\
 & -O3 & 106 & 106 & 106 & 106 &                               106 & 105 & 67 \\
 & -Os & 106 & 106 & 106 & 106 &                               106 & 106 & 78 \\
\hline                                                                          
\multirow{5}{*}{Clang 3.8} & -O0 & 106 & 106 & 106 & 106 &     106 & 106 & 93 \\
 & -O1 & 106 & 106 & 106 & 106 &                               106 & 106 & 90 \\
 & -O2 & 106 & 106 & 106 & 106 &                               106 & 106 & 85 \\
 & -O3 & 106 & 106 & 106 & 106 &                               106 & 106 & 85 \\
 & -Os & 106 & 106 & 106 & 106 &                               106 & 106 & 85 \\
\hline                                                                         
\multirow{5}{*}{Clang 6.0} & -O0 & 106 & 106 & 106 & 106 &    106 & 106 & 93 \\
 & -O1 & 106 & 106 & 106 & 106 &                              106 & 106 & 90 \\
 & -O2 & 106 & 106 & 106 & 106 &                              106 & 106 & 85 \\
 & -O3 & 106 & 106 & 106 & 106 &                              106 & 106 & 85 \\
 & -Os & 106 & 106 & 106 & 106 &                              106 & 106 & 84 \\
\hline
\multirow{5}{*}{ICC} & -O0 & 106 & 106 & 106 & 106 &       106 & 106 & 93 \\
 & -O1 & 106 & 106 & 106 & 106 &                           106 & 0 & 0 \\  
 & -O2 & 106 & 106 & 106 & 106 &                           93 & 0 & 0 \\   
 & -O3 & 106 & 106 & 106 & 106 &                           92 & 0 & 0 \\   
 & -Os & 106 & 106 & 106 & 106 &                           106 & 0 & 0 \\  
\hline
\multirow{5}{*}{GCC 9.2.1} & -O0 & 106 & 106 & 106 & 106 &    106 & 106 & 93 \\ 
 & -O1 & 106 & 106 & 106 & 106 &                          106 & 106 & 78 \\ 
 & -O2 & 106 & 106 & 106 & 106 &                          106 & 105 & 76 \\ 
 & -O3 & 106 & 106 & 106 & 106 &                          106 & 105 & 67 \\ 
 & -Os & 106 & 106 & 106 & 106 &                          106 & 106 & 78 \\ 
\hline                                                                      
\multirow{5}{*}{Clang 9.0.1} & -O0 & 106 & 106 & 106 & 106 &  106 & 106 & 93 \\ 
 & -O1 & 106 & 106 & 106 & 106 &                          106 & 106 & 89 \\ 
 & -O2 & 106 & 106 & 106 & 106 &                          106 & 106 & 85 \\ 
 & -O3 & 106 & 106 & 106 & 106 &                          106 & 105 & 84 \\ 
 & -Os & 106 & 106 & 106 & 106 &                          106 & 106 & 84 \\ 
\hline

  \end{tabular}
  \caption{Coreutils benchmarks functionality evaluation. Results broken down by compiler and optimization flag.}
  \label{table:functionality_detail_core}
\end{table*}

\begin{table*}[t]
  \centering
  \begin{tabular}{llr | rrr | rrr}
  \hline
  \multicolumn{9}{c}{CGC Benchmarks} \\ \hline
   \multirow{2}{*}{Compiler}& \multirow{2}{*}{Optimization} & \multirow{2}{*}{Binaries} & \multicolumn{3}{|c|}{\ddisasm{}} & \multicolumn{3}{|c}{\Ramblr{}}\\
   &  & &  Disasm  & Reassemble & Test  & Disasm & Reassemble & Test \\
    \hline
\multirow{6}{*}{GCC 5.5} & -O0 & 69 & 69 & 69 & 69 &           69 & 66 & 53 \\
 & -O1 & 68 & 68 & 68 & 68 &                               68 & 64 & 46 \\
 & -O2 & 69 & 69 & 69 & 69 &                               69 & 55 & 39 \\
 & -O3 & 68 & 68 & 68 & 68 &                               68 & 49 & 35 \\
 & -Ofast & 68 & 68 & 68 & 68 &                            68 & 49 & 35 \\
 & -Os & 68 & 68 & 68 & 68 &                               68 & 64 & 45 \\
\hline                                                                    
\multirow{6}{*}{GCC 7.1} & -O0 & 69 & 69 & 68 & 68 &         69 & 66 & 53 \\
 & -O1 & 69 & 69 & 69 & 69 &                               69 & 65 & 47 \\
 & -O2 & 68 & 68 & 68 & 67 &                               68 & 57 & 41 \\
 & -O3 & 69 & 69 & 69 & 68 &                               69 & 52 & 38 \\
 & -Ofast & 68 & 68 & 68 & 67 &                            68 & 52 & 37 \\
 & -Os & 68 & 68 & 68 & 68 &                               68 & 64 & 45 \\
\hline                                                                    
\multirow{6}{*}{Clang 3.8} & -O0 & 69 & 69 & 69 & 69 &         69 & 64 & 47 \\
 & -O1 & 69 & 69 & 69 & 68 &                               69 & 52 & 31 \\
 & -O2 & 68 & 68 & 68 & 68 &                               68 & 47 & 29 \\
 & -O3 & 68 & 68 & 68 & 68 &                               68 & 47 & 29 \\
 & -Ofast & 68 & 68 & 68 & 68 &                            68 & 47 & 29 \\
 & -Os & 69 & 69 & 69 & 69 &                               69 & 65 & 45 \\
\hline                                                                    
\multirow{6}{*}{Clang 6.0} & -O0 & 69 & 69 & 69 & 69 &     69 & 64 & 47 \\
 & -O1 & 69 & 69 & 69 & 69 &                               69 & 53 & 33 \\
 & -O2 & 69 & 69 & 69 & 69 &                               69 & 50 & 32 \\
 & -O3 & 68 & 68 & 68 & 68 &                               68 & 49 & 31 \\
 & -Ofast & 68 & 68 & 68 & 68 &                            68 & 49 & 31 \\
 & -Os & 68 & 68 & 68 & 68 &                               68 & 64 & 45 \\
\hline                                                                    
\multirow{6}{*}{ICC} & -O0 & 68 & 68 & 68 & 68 &           68 & 54 & 35 \\
 & -O1 & 69 & 69 & 69 & 68 &                               69 & 0 & 0 \\  
 & -Os & 68 & 68 & 68 & 67 &                               68 & 0 & 0 \\  
 & -O2 & 66 & 66 & 66 & 65 &                               66 & 0 & 0 \\  
 & -O3 & 66 & 66 & 66 & 64 &                               66 & 0 & 0 \\  
 & -Ofast & 65 & 65 & 65 & 63 &                            65 & 0 & 0 \\  
\hline                                                                    
\multirow{6}{*}{GCC 9.2.1} & -O0 & 69 & 69 & 68 & 68 &         69 & 68 & 55 \\
 & -O1 & 67 & 67 & 67 & 67 &                               67 & 66 & 48 \\
 & -O2 & 69 & 69 & 69 & 69 &                               69 & 58 & 41 \\
 & -O3 & 68 & 68 & 68 & 68 &                               68 & 55 & 39 \\
 & -Ofast & 69 & 69 & 69 & 69 &                            69 & 55 & 39 \\
 & -Os & 69 & 69 & 69 & 67 &                               67 & 66 & 45 \\
\hline                                                                    
\multirow{6}{*}{Clang 9.0.1} & -O0 & 69 & 69 & 69 & 69 &       69 & 66 & 49 \\
 & -O1 & 69 & 69 & 69 & 67 &                               69 & 54 & 33 \\
 & -O2 & 69 & 69 & 69 & 69 &                               69 & 52 & 33 \\
 & -O3 & 68 & 68 & 68 & 68 &                               68 & 50 & 32 \\
 & -Ofast & 68 & 68 & 68 & 68 &                            68 & 50 & 32 \\
 & -Os & 68 & 68 & 68 & 67 &                               68 & 65 & 44 \\
\hline
  \end{tabular}
  \caption{CGC benchmarks functionality evaluation. Results broken down by compiler and optimization flag.}
  \label{table:functionality_detail_cgc}
\end{table*}

\fi
\end{document}